\documentclass{iopart}

\usepackage{iopams} 
\usepackage{amstext}

\usepackage{graphicx}
\usepackage{dcolumn}
\usepackage{mathrsfs}
\usepackage{bm}
\usepackage{color,xspace}
\usepackage[small,compact,raggedright]{titlesec}
\usepackage{soul,hyperref}
\usepackage{iopams}
\usepackage{setstack}
\usepackage{graphicx}
\usepackage{xspace}

\newcommand{\eqref}[1]{{(\ref{#1})}}
\newcommand{\da}{{\Box_g}}
\newcommand{\daprime}{{\Box_{g'}}}

\newcommand{\ket}[1]{\left| {#1} \right\rangle}
\newcommand{\bra}[1]{\left\langle {#1} \right|}

\newcommand{\ii}{\mathrm{i}}
\newcommand{\ee}{\mathrm{e}}
\def\b{\begin{equation}}
\def\e{\end{equation}}

\begin{document}

\title[Cosmological quantum entanglement]{Cosmological quantum entanglement}

\author{Eduardo Mart\'in-Mart\'inez}

\address{Institute for Quantum Computing, Department of Physics and Astronomy, and Department of Applied Mathematics, University of Waterloo, 200 University
Avenue W, Waterloo, Ontario, N2L 3G1, Canada}

\author{Nicolas C. Menicucci}

\address{School of Physics, The University of Sydney, Sydney, NSW, 2006, Australia}

\begin{abstract}
We review recent literature on the connection between quantum entanglement and cosmology, with an emphasis on the context of expanding universes. We discuss recent theoretical results reporting on the production of entanglement in quantum fields due to the expansion of the underlying spacetime. We explore how these results are affected by the statistics of the field (bosonic or fermionic), the type of expansion (de~Sitter or asymptotically stationary), and the coupling to spacetime curvature (conformal or minimal). We then consider the extraction of entanglement from a quantum field by coupling to local detectors and how this procedure can be used to distinguish curvature from heating by their entanglement signature. We review the role played by quantum fluctuations in the early universe in nucleating the formation of galaxies and other cosmic structures through their conversion into classical density anisotropies during and after inflation. We report on current literature attempting to account for this transition in a rigorous way and discuss the importance of entanglement and decoherence in this process. We conclude with some prospects for further theoretical and experimental research in this area. These include extensions of current theoretical efforts, possible future observational pursuits, and experimental analogues that emulate these cosmic effects in a laboratory setting.
\end{abstract}


\maketitle
\section{Introduction}

The study of the early universe and the large-scale structure of the spacetime has flourished since the introduction of the theory of inflation~\cite{Guth1981}, tested by the observation of anisotropies in the cosmic microwave background~\cite{Smoot1992, Peiris2003}. Many questions still remain, however. In the late 90s it was found that our universe undergoes an accelerated expansion~\cite{Riess1998,Perlmutter1999}, implying that most of the content of the universe is some form of dark energy whose nature remains unknown. Also, measurements of the deuterium-hydrogen ratio in interstellar absorption, together with some theoretical models of cosmological nucleosynthesis, have provided evidence that ordinary baryonic matter is responsible for only about 20\% of the matter content of the universe and that there should be some sort of non-baryonic cold dark matter~\cite{Reid2010} accounting for the rest. The abundance of this dark matter cannot be explained exclusively with standard-model neutrinos~\cite{neutrinoref}. This mysterious dark matter remains elusive even to the latest progress in experimental particle physics in the Large Hadron Collider~\cite{LHC1,LHC2,LHC3,LHC4}.

Another important ingredient of a consistent cosmological model is still missing:\ we do not fully understand the nature of the quantum field responsible for the hypothetical inflation period in the very early universe~\cite{Peiris2003}. In the absence of a full quantum theory for gravity, which would include backreaction of the quantum fields on the curvature of spacetime, quantum field theory in curved spacetime (without backreaction) is the most complete theory so far~\cite{Birrell1982}. The gravitational curvature of spacetime has nontrivial effects on quantum fields living on the spacetime when compared with their flat-spacetime counterparts. This is especially interesting in the case of dynamical spacetime backgrounds and/or when the spacetime includes horizons preventing an observer from having experimental access to the full quantum state of a field~\cite{GibHawking}. It is known that the gravitational interaction may induce or reveal inaccessible quantum correlations in the field state in scenarios such as expanding universes or stelar collapse~\cite{caball, fermxpanding, colapse} and that particle detectors interacting with a quantum field can become entangled~\cite{Reznik2005} in a way that is sensitive to the structure of the spacetime that those fields inhabit~\cite{VerSteeg2009, Nambu2011}.

As we will discuss in this paper, the early universe's energy content was dominated by a highly entangled quantum field background. If the quantum correlations managed to survive to our days in weakly interacting fields in some form, they can provide precise information about the nature and history of the spacetime~\cite{Nambu2008, Simon2000, Campo2005}. Their study may prove useful in constructing models of the early universe and in gaining information about the phenomenon of inflation. In particular, we would like to understand the decoherence mechanisms that governed the transition from a highly non-classical quantum-correlated field state to the current classical background. By means of this understanding we may be able to provide answers to the still unsolved questions about the inflationary theory.

In this article, we review the different scenarios studied in the literature in which quantum entanglement emerges in the study of quantum fields in an expanding spacetime. Using the tools of quantum information science, we analyse the appearance of quantum correlations in quantum fields due to gravitational expansion of the underlying spacetime, paying attention to the role played by such correlations in the current cosmological models. We discuss whether these quantum correlations may be present today as a remanent of the primitive universe. We also discuss the possibility of extracting this entanglement through local interaction with particle detectors and consider how such experiments could provide information about the cosmological parameters of the universe. We conclude with a discussion of possible applications and future research avenues in this direction. Natural units are used throughout:\ $\hbar = G = c = k_B = 1$.

\section{Introduction to particle creation in expanding universes}
\label{sec:particlecreation}

Let us consider a universe undergoing a homogeneous and isotropic expansion. This kind of spacetime is very well described in terms of  the well-known Friedmann-Lema\^itre-Robertson-Walker~(FLRW) metric, 
\begin{equation}\label{FLRWm1}
ds^2=dt^2-[a(t)]^2 d\Sigma^2\,,
\end{equation}
where $d\Sigma$ is an element of a 3-dimensional space of uniform curvature, either elliptical, hyperbolic, or Euclidean. Expanding $d\Sigma$ gives
\begin{equation}
ds^2=dt^2-[a(t)]^2 \left(\frac{dr^2}{1-kr^2}+r^2d\Omega^2\right)\,,
\end{equation}
where $d\Omega$ is the line element in the unit sphere and $k$ characterises the curvature of the space (negative, zero, or positive for hyperbolic, Euclidean, or elliptical, respectively). In the case where the spatial geometry is Euclidean ($k=0$) we can write this metric as
\begin{equation}\label{FLRWm}
ds^2=dt^2-[a(t)]^2 (dx^2+dy^2+dz^2)\,.
\end{equation}
Even if the spatial geometry of the universe is not exactly Euclidean, the spatial curvature is close enough to zero to place the radius at approximately the horizon of the observable universe or beyond, so this consideration seems reasonable, and we will admit it from now on.

\subsection{Scalar field quantisation in general spacetimes}
\label{sec:scalarfieldq}

Before considering quantum fields living in arbitrary curved spacetimes, let us quickly review for the reader the well-known process of field quantisation in flat spacetime. This will serve a double purpose:\ to introduce the notation that we will use in following sections and to highlight some aspects of the process of field quantisation that become relevant when we apply it to arbitrary backgrounds. Let us consider an inertial observer in flat spacetime whose proper coordinates are the global Minkowski coordinates~$(t,x, y, z)$. He wants to build a quantum field theory for a free masless scalar field~$\Phi$ whose equation of motion is 
\begin{equation}\label{KG1}
(\Box +m^2)\Phi =0\,.
\end{equation}
Since Minkowski spacetime admits a global timelike Killing vector~$\partial_t$, the solutions to this equation can be classified into positive and negative frequency. A positive frequency solution of Eq.~\eqref{KG1}, $u_k(\bm x,t)$, therefore satisfies
\begin{equation}\label{flatcriterion}
\partial_t u_k(\bm x,t)=-\ii\omega_ku_k(\bm x,t)\,.
\end{equation}
Importantly, this criterion would be the same if, instead of $t$, we consider the proper time of any inertial observer. Treating $\Phi$ as a classical field for the moment, we can expand an arbitrary solution $\Phi(\bm x,t)$ of Eq.~\eqref{KG1} as a sum of these positive-frequency and negative-frequency solutions with $c$-number coefficients. Hence, we can express  $\Phi(\bm x,t)$ as a combination of positive-frequency $u_i(\bm x,t)$ and negative-frequency $u^*_i(\bm x,t)$ solutions of Eq.~\eqref{KG1}:
\begin{equation}\label{exp1}
\Phi(\bm x,t)=\sum_i \left[\alpha_i u_i(\bm x,t)+\alpha_i^* u^*_i(\bm x,t)\right]\,.
\end{equation}
This definition will agree with the definition for any arbitrary inertial observer.\footnote{For notational convenience, we are using the sum symbol~$\sum_i$ even when integration over frequencies is meant. In this case, $\sum_i \mapsto \int d^{D-1}k$, where $d^{D-1}k$ is the relevant integration measure.}

The solutions $u_i(\bm x,t)$ (and their complex conjugates) form an orthonormal basis of solutions with respect to the Klein-Gordon scalar product, defined through the continuity equation to be
\begin{equation}\label{KGsc}
(u_j,u_k)=-\ii\int \text{d}^3x\left(u_j\partial_tu_k^*-u_k^*\partial_tu_j\right)\,,
\end{equation}
which is positive definite over the space of positive energy solutions and satisfies
\begin{equation}\label{KGproperties}
(u_1,u_2)=(u_2,u_1)^*\,, \qquad (u_1^*,u_2^*)=-(u_2,u_1)\,.
\end{equation}
The modes $u_j$ fulfil the orthonormality relations
\begin{equation}\label{orto}
(u_j,u_k)=\delta_{jk}=-(u_j^*,u_k^*)\,, \qquad (u_j,u_k^*)=0\,.
\end{equation}
We can now construct a Fock space for the corresponding quantum field by following the standard canonical field quantisation scheme. We promote the classical Klein-Gordon field to a quantum field operator satisfying the standard equal-time commutation relations
\begin{eqnarray}
	\left[\Phi(\bm x,t),\Pi(\bm x',t)\right] &=\ii\delta(\bm x-\bm x')\,, \nonumber \\
	\left[\Phi(\bm x,t),\Phi(\bm x',t)\right] &=\left[\Pi(\bm x,t),\Pi(\bm x',t)\right]=0\,,
\end{eqnarray}
where $\Pi(\bm x,t)=\partial_t \Phi(\bm x,t)$ is the canonical momentum associated with the field operator $\Phi$. This means that we replace the complex amplitudes $\alpha_i$ and $\alpha_i^*$ by annihilation and creation operators $a_i$ and $a_i^\dagger$, which inherit the standard commutation relations $[a_i,a^\dagger_j]=(u_i,u_j)=\delta_{ij}$, $[a_i,a_j]=[a^\dagger_i,a^\dagger_j]=0.$ We now construct the Fock space:\ first we characterise the vacuum state of the field  as the state which is annihilated by all the $a_i$:
\begin{equation}
a_i\ket{0}=0 \quad \forall i\,.
\end{equation}
Then we define the one-particle Hilbert space by applying the creation operators $a_i^\dagger$ to the vacuum state,
\begin{equation}
\ket{1_i}=a^\dagger_i\ket{0}\,,
\end{equation}
and by repeatedly applying the particle creator we build the complete Fock space:
\begin{equation}
\ket{n^1_{i_1},n^2_{i_2},\dots,n^k_{i_k}}=\frac{1}{\sqrt{n^1!n^2!\dots n^k!}} (a^\dagger_{i_1})^{n^1}(a^\dagger_{i_2})^{n^2}\dots(a^\dagger_{i_k})^{n^k}\ket{0}\,.
\end{equation}

Due to the symmetries of Minkowski spacetime, this quantisation procedure is independent of the particular choice of the inertial observer. Any other choice of time $t$ is related to this one via Poincar\'e transformations which do not modify what we would define as positive and negative frequency modes. As a consequence, the field expansion Eq.~\eqref{exp1} can be performed equivalently for any inertial reference frame, and the splitting between positive and negative frequency modes is invariant. Hence, the vacuum state is also Poincar\'e invariant, and the construction of the Fock space is equivalent for any inertial observer. For general spacetimes, however, we cannot assume that we have global Poincar\'e symmetry, and thus we lose the objective way to build quantum fields. Instead, the vacuum will not be unique, and the notion of ``particle'' necessarily becomes observer dependent.

If we continue analysing the scalar field but in a general background, we have to substitute the D'Alambertian in Eq.~\eqref{KG1} with the covariant D'Alambert operator, which we will denote as $\da$ and which has the form
\begin{equation}
\da=\frac{1}{\sqrt{|g|}}\partial_\mu \sqrt{|g|} g^{\mu\nu}\partial_\nu\,,
\end{equation}
so that Eq.~\eqref{KG1} now reads, quite simply,
\begin{equation}\label{KG2}
(\da + m^2)\Phi  =0\,.
\end{equation}
Here, we have assumed that there is no coupling of the field with the spacetime scalar curvature. This is called \emph{minimal coupling}. But Eq.~\eqref{KG2} is not the most general equation for a scalar field in a curved spacetime background. In fact, there are some interesting aspects of the scalar field equation in curved backgrounds with more general couplings (especially, conformal coupling) that will be dealt with in Sec.~\ref{conformal}, but we will keep the equation in its simplest form for now. To extend the Klein-Gordon product, Eq.~\eqref{KGsc}, to curved spacetime, we need a complete set of initial data---in other words, a Cauchy hypersurface~$\Sigma$ over which we can extend the integral:\footnote{Note that Gauss's theorem guarantees that the product is independent of the choice of the Cauchy hypersurface~$\Sigma$.}
 \begin{equation}\label{KGscc}
(u_j,u_k)=-\ii\int \text{d}\Sigma\, n^\mu\left(u_j\partial_\mu u_k^*-u_k^*\partial_\mu u_j\right)\,,
 \end{equation}
where $d\Sigma$ is the volume element, and $n^\mu$ is a future-directed timelike unit vector orthogonal to~$\Sigma$.

The operative question is whether the spacetime is stationary or not. If it is, then it has a timelike Killing vector field~$\xi^\mu$, and we have a natural way to define positive frequency modes~($u_j$) in an analogous way as we did for flat spacetime in Eq.~\eqref{flatcriterion}:
\begin{equation}\label{criterion2}
\xi^\mu\nabla_\mu u_j = -\ii \omega_j u_j\,,
\end{equation}
where $\omega_j>0$ and $\nabla_\mu$ is the covariant derivative. Of course, we can construct a local set of coordinates whose timelike coordinate is the Killing time~$\tau$ associated to the isometry~$\xi^\mu$ such that it satisfies $\xi^\mu\nabla_\mu \tau =1$. For flat spacetime, which is a particular case of stationary spacetime, the role of $\tau$ is played by the coordinate $t$. In contrast, non-stationary spacetimes do not admit a global timelike Killing vector field, and therefore there is no natural way to uniquely distinguish positive- and negative-frequency solutions of the field equation. In the absence of this metric symmetry there is an ambiguity when it comes to defining particle states. Without a natural way to split modes into positive and negative frequencies, there is no objective way to construct a unique Fock space, starting from the fact that there is no unique notion of a vacuum state. This means that there is no unique, observer-independent notion of the particle content of a quantum field on a non-stationary background. Nevertheless, we can still find an approximate particle interpretation when the spacetime posses asymptotically stationary regions, as we will see in the following section.

\subsection{Bogoliubov transformations in asymptotically stationary spacetimes}\label{nonstaint}

There are some scenarios in which the spacetime is not stationary but possesses stationary asymptotic regions. This is the case, for example, of some models of expansion of the universe \cite{dun1} or the stellar collapse and formation of black holes \cite{NavarroSalas}. 

Consider a spacetime that has asymptotically stationary regions in the past and in the future. We will denote them `in' and `out' respectively. In these regions we can give a particle interpretation to the solutions of the field equations. Namely, we can build two different sets of solutions of the field, the first one~$\{u_{\hat{\bm{k}}_j}^{\text{in}}\}$ made of modes with positive frequency~$\hat\omega_j$ with respect to the comoving time in the asymptotic past. The second set~$\{u_{{\bm{k}}_j}^{\text{out}}\}$ consisting of modes with positive frequency~$\omega_j$ with respect to the comoving time in the future. In this fashion we can now expand the quantum field in terms of both sets of solutions to the field equation
\begin{equation}\label{minkexp1}
\Phi=\sum_i \left(a_{\hat{\bm{k}}_i,\text{in}}u_{\hat{\bm{k}}_i}^{\text{in}}+a_{\hat{\bm{k}}_i,\text{in}}^\dagger u_{\hat{\bm{k}}_i}^{\text{in}*}\right)=\sum_i \left(a_{{\bm{k}}_i,\text{out}}u_{{\bm{k}}_i}^{\text{out}}+a_{{\bm{k}}_i,\text{out}}^\dagger u_{{\bm{k}}_i}^{\text{out}*}\right)\,.
\end{equation}
What is more, since both set of modes are complete, one can also expand one set of modes in terms of the other by means of the Klein-Gordon scalar product:
\begin{eqnarray}\label{modinout1}
u^{\text{in}}_{\hat{\bm{k}}_i}&= \sum_j \left[(u_{\hat{\bm{k}}_i}^{\text{in}},u_{{\bm{k}}_j}^{\text{out}})u_{{\bm{k}}_j}^{\text{out}}-(u_{\hat{\bm{k}}_i}^{\text{in}},u_{{\bm{k}}_j}^{\text{out}*})u_{{\bm{k}}_j}^{\text{out}*}\right]\,,\\
u^{\text{out}}_{{\bm{k}}_i}&= \sum_j \left[(u_{{\bm{k}}_i}^{\text{out}},u_{\hat{\bm{k}}_j}^{\text{in}})u_{\hat{\bm{k}}_j}^{\text{in}}-(u_{{\bm{k}}_i}^{\text{out}},u_{\hat{\bm{k}}_j}^{\text{in}*})u_{\hat{\bm{k}}_j}^{\text{in}*}\right]\,.
\end{eqnarray}
We define the Bogoliubov coefficients as
\begin{equation}
\alpha_{ij}=(u_{{\bm{k}}_i}^{\text{out}},u_{\hat{\bm{k}}_j}^{\text{in}})\,,\qquad \beta_{ij}=-(u_{{\bm{k}}_i}^{\text{out}},u_{\hat{\bm{k}}_j}^{\text{in}*})\,.
\end{equation}
Then, using the properties of the Klein-Gordon product, Eq.~\eqref{KGproperties}, we can rewrite Eq.~\eqref{modinout1} as
\begin{eqnarray}\label{modinout2}
u^{\text{in}}_{\hat{\bm{k}}_i}&= \sum_j \left[\alpha_{ji}^{*}u_{{\bm{k}}_j}^{\text{out}}-\beta_{ji}u_{{\bm{k}}_j}^{\text{out}*}\right]\,,\\
u^{\text{out}}_{{\bm{k}}_i}&= \sum_j \left[\alpha_{ij}u_{\hat{\bm{k}}_j}^{\text{in}}+\beta_{ij}u_{\hat{\bm{k}}_j}^{\text{in}*}\right]\,.
\end{eqnarray}
Now, we can expand the particle operators associated with one basis in terms of operators of the other basis
\begin{eqnarray}
\label{uno1}a_{\hat{\bm{k}}_i,\text{in}}&=\sum_j \left(\alpha_{ji} a_{{\bm{k}}_j,\text{out}}+\beta_{ji}^* a_{{\bm{k}}_j,\text{out}}^\dagger\right),\\
\label{uno2}a_{{\bm{k}}_i,\text{out}}&=\sum_j \left(\alpha^*_{ij} a_{\hat{\bm{k}}_j,\text{in}}-\beta_{ij}^* a_{\hat{\bm{k}}_j,\text{in}}^\dagger\right).
\end{eqnarray}

Now let us consider the vacuum state in the asymptotic past region $\ket0_{\text{in}}$, which is annihilated by $a_{\hat{\bm{k}}_i,\text{in}}\ \forall\hat\omega_i$. We would like to know the form of the state $\ket0_{\text{in}}$ in the basis of solutions of the Klein-Gordon equation in the asymptotic future. To compute this we use the fact that  $a_{\hat{\bm{k}}_i,\text{in}}\ket0_{\text{in}}=0$. If we substitute $a_{\hat{\bm{k}}_i,\text{in}}$ in terms of `out' operators using Eq.~\eqref{uno1}, we obtain
\begin{equation}\label{condit}
\sum_j \left(\alpha_{ji} a_{{\bm{k}}_j,\text{out}}+\beta_{ji}^* a_{{\bm{k}}_j,\text{out}}^\dagger\right)\ket{0}_{\text{in}}=0\,.
\end{equation}
Without loss of generality, we now make an ansatz for the general form of the state $\ket0_{\text{in}}$ in terms of the `out' Fock basis as a sum of its $n$-particle amplitudes
\begin{equation}\label{ansatz1}
\ket{0}_{\text{in}}= C\ket{0}_{\text{out}}+C^{j_1}\ket{\Psi}_{j_1}+C^{j_1,j_2}\ket{\Psi}_{j_1,j_2}+\dots+C^{j_1,\dots,j_n}\ket{\Psi}_{j_1,\dots,j_n}+\dots\,,
\end{equation}
where each summand has the form
\begin{equation}
C^{j_1,\dots,j_n} \ket{\Psi_n}_{j_1,\dots,j_n}=\sum_{j_1,\dots,j_n} C^{j_1,\dots,j_n} a_{{\bm{k}}_{j_1},\text{out}}^\dagger\dots a_{{\bm{k}}_{j_n},\text{out}}^\dagger\ket{0}_{\text{out}}\,.
 \end{equation}
 Substituting Eq.~\eqref{ansatz1} in Eq.~\eqref{condit}, we obtain an infinite number of constraints for the coefficients $C^{j_1,\dots,j_n}$.  We can readily see that the zero-particle component of Eq.~\eqref{condit}  can only be obtained from the annihilator acting on the 1-particle component of the state defined in Eq.~\eqref{ansatz1}, thus giving the condition
\begin{equation}\label{coefcond1}
\sum_j\alpha_{j i}C^j=0 \quad \Longrightarrow \quad C^j= 0\,.
\end{equation} 
Now, the $n$-particle component ($n\neq0$) of Eq.~\eqref{condit} is obtained by acting with the annihilator on the ($n+1$)-particle component of Eq.~\eqref{ansatz1} and with the creator on the ($n-1$)-particle component of Eq.~\eqref{ansatz1}. Thus, we know that the coefficients $C^{j_1,\dots,j_{n+1}}$ can be written as functions of the coefficients $C^{j_1,\dots,j_{n-1}}$, providing a recurrence relation that can be used to write all the coefficients as functions of $C$ (the vacuum coefficient).

This, together with Eq.~\eqref{coefcond1}, means that the `in' vacuum evolves to a state in the asymptotic future with no odd number of particles components, and that the coefficients of even components in Eq.~\eqref{ansatz1} are related pairwise. To find the form of this coefficients we use the inverse of the Bogoliubov coefficient matrix $\alpha_{ij}$, which is a matrix whose $ij$ component is denoted~$(\alpha^{-1})_{ij}$. Therefore, given arbitrary $C_j$ and $D_j$, the following identity holds:
\begin{equation}\label{inve}
\sum_j\left(\alpha_{j i}C_j+\beta_{ji}^*D_j\right)=0 \qquad \Longrightarrow \qquad C_k=-\sum_{ij}\beta^*_{ji}(\alpha^{-1})_{ ik}D_j\,.
\end{equation}
After some basic but lengthy algebra, we obtain the following expression for the `in' vacuum in terms of `out' modes:
\begin{equation}\label{dynoresult}
\ket{0}_{\text{in}}=C\exp\left(-\frac12\sum_{ijk}\beta^*_{ik}(\alpha^{-1})_{kj}a_{{\bm{k}}_{i},\text{out}}^\dagger a_{{\bm{k}}_{j},\text{out}}^\dagger\right)\ket{0}_{\text{out}}\,.
\end{equation}
This is a Gaussian state, with $C$ obtained by imposing normalisation. Notice that this state is, in general, not separable. This means that, depending on $\alpha_{ij}$ and $\beta_{ij}$, the final state of the field can possess nontrivial quantum correlations.

If we now compute the expectation value of the number operator in the asymptotic future $N^{\text{out}}_{{\bm{k}}_{j}}=a_{{\bm{k}}_{j},\text{out}}^\dagger a_{{\bm{k}}_{j},\text{out}}$ when the state of the field is the vacuum in the asymptotic past,  we find that 
\begin{equation}\label{production}
\langle N^{\text{out}}_{{\bm{k}}_{j}}\rangle_{\text{in}} = {}_{\text{in}}\!\bra{0}N^{\text{out}}_{{\bm{k}}_{j}}\ket{0}_{\text{in}}=\sum_i \left|\beta_{ij}\right|^2.
\end{equation}
This implies that if $\beta_{ij}$ is different from zero, one would observe particle production as a consequence of the expansion.

\subsection{FLRW Universe I: Particle production in a fast expansion period}\label{fastexp}

We have discussed in Sec.~\ref{nonstaint} that when a non-stationary spacetime possesses asymptotically stationary regions, it is possible to relate field modes in the asymptotic past to modes in the asymptotic future, being able to compute the field quanta creation due to the gravitational interaction. This is the case for the FLRW spacetime,
 Eq.~\eqref{FLRWm}, if we impose the additional conditions
\begin{equation}\label{condasy}
a(+\infty)\rightarrow \text{const.}\,, \qquad  a(-\infty)\rightarrow \text{const.}\,,
\end{equation}
with the former constant larger than the latter. Although measurements of distant supernovae have demonstrated that our universe is undergoing an accelerating expansion, to good degree of approximation, certain models fulfilling Eq.~\eqref{condasy} (as the one that we will study below) can very well describe some inflationary scenarios. Roughly speaking, these include a period of slow expansion, then a period of very fast expansion, and then again a period of slow expansion.

As can be seen from Eq.~\eqref{uno2}, computing the analytic form of the particle operators in the asymptotic future in terms of the operators in the asymptotic past is not straightforward. There are, however, some particular solvable toy models that allow for the study of fundamental phenomena. These include scalar models~\cite{Birrell} and spin-$\frac 1 2$ fermionic models~\cite{dun1}. To illustrate this, let us study the scalar field case within an exactly solvable model in 1+1 dimensions.

Let us consider the FLRW metric Eq.~\eqref{FLRWm} and rewrite it in terms of the conformal time coordinate
\begin{equation}\label{conFLRW}
\eta=\int_0^t \frac{\text{d}\tau}{a(\tau)}\,.
\end{equation}
In doing this, we obtain
\begin{equation}\label{MconFLRW}
ds^2=[a(\eta)]^2(d\eta^2- dx^2)\,.
\end{equation}
Following the standard notation, we define $C(\eta)=[a(\eta)]^2$. Let us assume the following specific form for the conformal factor:
\begin{equation}\label{factorbos}
C(\eta)=1+\epsilon \tanh(\rho\eta)\,,
\end{equation}
where $\epsilon$ and $\rho$ are positive real parameters controlling the total volume and
rapidity of the expansion, respectively. Imposing this specific form for the conformal factor makes it simple to study analytically while at the same time allowing us to study the fundamental behaviour as a function of the rapidity of the expansion and its total volume.

It is interesting to note that for large $\epsilon$ the FLRW metric with this conformal factor behaves, for $0<t\ll \rho^{-1}$, like the radiation-dominated Friedman universe, with an exponentially fast approach to asymptotic flatness for $t\ll \rho^{-1}$. As mentioned above, we can consider this as a convenient way to approximate a spacetime that is asymptotically flat in the past and future but undergoes rapid inflation in between (with smooth transitions). The asymptotic flatness at each end allows us to meaningfully define different but physically meaningful particle states for the asymptotic past and future. Note that this example is analytically solvable only for the scalar field. The same $C(\eta)$ does not provide an analytically solvable model in the fermionic case, which is discussed in Sec.~\ref{DiracF}.

The Klein-Gordon equation in the asymptotic past and the asymptotic future in this particular case admits solutions of the form
\begin{eqnarray}
u^{\text{in}}_k(x,\eta)&=\chi^{\text{in}}(\eta) \exp\left[\ii (kx-\omega_{_+}\eta)-\frac{\ii\omega_-}{\rho}\ln[2\cosh(\rho\eta)] \right]\,, \\
u^{\text{out}}_k(x,\eta)&=\chi^{\text{out}}(\eta) \exp \left[\ii (kx-\omega_{_+}\eta)-\frac{\ii\omega_-}{\rho}\ln[2\cosh(\rho\eta)] \right]\,,
\end{eqnarray}
where
\begin{eqnarray}
\chi^{\text{in}}(\eta)&=\frac{1}{2\sqrt{\pi \omega_{\text{in}}}}F_1\left[1+\frac{\ii\omega_-}{\rho},\frac{\ii\omega_-}{\rho};1-\frac{\ii\omega_{\text{in}}}{\rho};\frac{1}{2}[1-\tanh(\rho\eta)]\right]\,, \\
\chi^{\text{out}}(\eta)&=\frac{1}{2\sqrt{\pi \omega_{\text{out}}}}F_1\left[1+\frac{\ii\omega_-}{\rho},\frac{\ii\omega_-}{\rho};1-\frac{\ii\omega_{\text{out}}}{\rho};\frac{1}{2}[1-\tanh(\rho\eta)]\right]\,,
\end{eqnarray}
where the $F_1$ are hypergeometric functions, and
\begin{equation}
\omega_{\text{out/in}}=\sqrt{k^2+m^2(1\pm\epsilon)}\,, \qquad
\omega_{\pm}=\frac12(\omega_{\text{out}}\pm \omega_{\text{in}})\,.
\end{equation}
It can be checked that indeed
\begin{eqnarray}
u^{\text{in}}_k(x,\eta)&\underset{\eta\rightarrow-\infty}{\longrightarrow}\frac{1}{2\sqrt{\pi \omega_{\text{in}}}}\ee^{\ii (kx-\omega_{\text{in}}\eta)},\\
u^{\text{out}}_k(x,\eta)&\underset{\eta\rightarrow\infty}{\longrightarrow}\frac{1}{2\sqrt{\pi \omega_{\text{out}}}}\ee^{\ii (kx-\omega_{\text{out}}\eta)}.
\end{eqnarray}
Using the properties of the hypergeometric functions, the computation of the Bogoliubov coefficients matrices take a diagonal form in this case $\alpha_{kj}=\alpha^*_k\delta_{kj}$, $\beta_{kj}=-\beta_k\delta_{kj}$, so that in Eq.~\eqref{modinout2} the Bogoliubov transformations only mix modes of the same~$k$:
\begin{equation}\label{modolibov}
u^{\text{in}}_k(x,\eta)=\alpha_k  u^{\text{out}}_k(x,\eta)+\beta_k u^{{\text{in}*}}_{-k}(x,\eta)\,,
\end{equation}
where
\begin{eqnarray}\label{alphak}
\alpha_k&=\sqrt{\frac{\omega_{\text{out}}}{\omega_{\text{in}}}}\frac{\Gamma([1-(\ii\omega_{\text{in}}/\rho)])\Gamma(-\ii\omega_{\text{out}}/\rho)}{\Gamma([1-(\ii\omega_{+}/\rho)])\Gamma(-\ii\omega_{+}/\rho)}\,, \\[3mm]
\label{betak}\beta_k&=\sqrt{\frac{\omega_{\text{out}}}{\omega_{\text{in}}}}\frac{\Gamma([1-(\ii\omega_{\text{in}}/\rho)])\Gamma(\ii\omega_{\text{out}}/\rho)}{\Gamma([1+(\ii\omega_{-}/\rho)])\Gamma(\ii\omega_{-}/\rho)}\,.
\end{eqnarray}
It is important to be clear why the Bogoliubov transformation only mixes modes of the same momentum. This fact is not related to the particular expansion model choice in Eq.~\eqref{factorbos}, nor to the dimension of the spacetime considered. Rather, it is a consequence only of the conformal symmetry (up to mass terms in the minimal coupling scenario) of the theory and the conformal flatness of the metric. This conformal equivalence relates the equation of motion in the FLRW scenario to its flat-spacetime form. This means that the general form of the Bogoliubov transformation Eq.~\eqref{modolibov} is valid for any FLRW universe with sufficiently well-behaved scaling factor and minimally coupled field, where Eq.~\eqref{condasy} is fulfilled \cite{dun1}.

From Eq.~\eqref{production} we see that the mean number of particles in the asymptotic future when the field was prepared in the vacuum state in the past is
\begin{equation}
\langle N^{\text{out}}_{k}\rangle_{\text{in}}=|\beta_k|^2,
\end{equation}
That can be interpreted as the creation of quanta in the mode $k$ of the field as a consequence of the spacetime expansion. Notice that when $m=0$, $\omega_{\text{in}}=\omega_{\text{out}}$, which means that $|\beta_k|=0$, and no particles are present in the asymptotic future. This is because, in the particular case of two spacetime dimensions, for $m=0$, the theory is also conformally invariant. Let us study this case in more detail.

\subsection{FLRW Universe II: Particle detection in the conformal vacuum}
\label{conformal}

So far we have only considered the case where the field is not coupled to the scalar curvature (Ricci scalar) of the spacetime. This is known as minimal coupling to curvature. Indeed, Eq.~\eqref{KG2} comes from a particular case of the more general action associated with a scalar field in a curved spacetime background of dimension~$D$, which is
\begin{equation}\label{action}
\mathcal{S}=\int \text{d}^D x \frac{\sqrt{|g|}}{2}\left[g^{\mu\nu}\partial_\mu \Phi\partial_\nu \Phi-(m^2+\xi R)\Phi^2 \right]\,,
\end{equation}
whose associated equation of motion is
\begin{equation}\label{field}
\left(\da + m^2+\xi R \right)\Phi=0\,,
\end{equation}
where $\xi$ is the coupling of the field to the Ricci curvature scalar~$R$. In the section above, we have considered the case where $\xi=0$ (a.k.a. minimal coupling). However, there is another case of great interest in cosmological scenarios given that the FLRW metrics are conformally flat. This case is that of \emph{conformal coupling.}

Consider what happens if we make a conformal transformation of the metric:
\begin{equation}
g'_{\mu\nu}=[\Omega(x)]^2g_{\mu\nu}\,.
\end{equation}
The geometrical quantities in Eq.~\eqref{action} are then changed as follows:
\begin{eqnarray}
{g'}^{\mu\nu}&=[\Omega(x)]^{-2}g_{\mu\nu}\,, \\ 
\sqrt{|g'|}&=[\Omega(x)]^D\sqrt{|g|}\,, \\
R'&=\Omega^{-2}\left[R-2(D-1)\da \ln\Omega+(1-D)(D-2)g^{\eta\gamma}(\ln \Omega)_{,\eta}(\ln \Omega)_{,\gamma}\right]\,.
\end{eqnarray}
Through simple dimensional analysis we find that the field has dimensions of $[\text{length}]^{\frac{2-D}{2}}$ and hence it is also rescaled under a conformal transformation:
\begin{equation}
\Phi'=\Omega^{(2-D)/2}\Phi.
\end{equation}
Now, if we want the action in Eq.~\eqref{action} to be invariant under this conformal transformation (except for a boundary term), we find that there are three contributions spoiling such invariance: the mass term, the derivatives in the kinetic term (if $\Omega$ is not constant), and the coupling to the curvature term. If we look at the massless case, however, we can always choose the coupling to the curvature such that the action becomes conformally invariant---namely,
\begin{equation}
\xi=\frac{(D-2)}{4(D-1)}\,.
\end{equation}
This case (with $m=0$) is called conformal coupling. In two spacetime dimensions, we see that if $m=0$ the minimal coupling $\xi=0$ gives conformal invariance of the action, so it coincides exactly with the conformal coupling. In contrast, in four spacetime dimensions, we require $m=0$  and $\xi=1/6$ to have a conformaly invariant action. This is relevant to our interests since, if we apply the conformal transformation to the field equation Eq.~\eqref{field} with $D=4$, $m=0$, and $\xi=1/6$ (conformal coupling), then we obtain
\begin{equation}\label{truci}
\left(\da+\frac16 R\right)\Phi=\Omega^3 \left(\daprime+\frac16 R'\right)\Phi'\,.
\end{equation}
Hence, if $\Phi'$ is a solution of the conformally transformed equation, then $\Phi=\Omega\Phi'$ will be a solution of the original equation.

Now, if we consider the FLRW metric from Eq.~\eqref{FLRWm} and rewrite it in terms of the conformal time coordinate
\begin{equation}
\eta=\int_0^t \frac{\text{d}\tau}{a(\tau)}\,,
\end{equation}
we obtain
\begin{equation}
ds^2=[a(\eta)]^2(d\eta^2- dx^2-dy^2-dz^2)\,.
\end{equation}
A conformal transformation $\Omega=[a(\eta)]^{-1}$ transforms this metric into the Minkowski metric so that $g'_{\mu\nu}=\eta_{\mu\nu}$. For $\eta_{\mu\nu}$, $R'=0$ and the field equation is just the usual wave equation, which admits solutions of the well-known form
\begin{equation}
v_{\bm k}\propto \exp[\ii(\bm k \cdot \bm x- |\bm k| \eta)]=\exp\left[\ii\left(\bm k \cdot \bm x- \int_0^t \omega(\tau)\, \text{d}\tau \right)\right]\,,
\end{equation}
where we have defined $\omega(t)=|\bm k|/a(t)$. Using Eq.~\eqref{truci}, we have that the original equation admits solutions of the form~$u_{\bm k}=[a(\eta)]^{-1}v_{\bm k}$. Now, $u_{\bm k}$ are positive frequency solutions at early times, so the field admits the expansion
\begin{equation}
\Phi=\sum_i \left(a_{\hat{\bm{k}}_i,\text{in}}u_{\hat{\bm{k}}_i}+a_{\hat{\bm{k}}_i,\text{in}}^\dagger u_{\hat{\bm{k}}_i}^{*}\right)\,.
\end{equation}
But $u_{\bm k}$ are also positive-frequency solutions at late times, so the field also admits the expansion
\begin{equation}
\Phi=\sum_i \left(a_{{\bm{k}}_i,\text{out}}u_{{\bm{k}}_i}+a_{{\bm{k}}_i,\text{out}}^\dagger u_{{\bm{k}}_i}^{*}\right)\,.
\end{equation}
In this case, the Bogoliubov transformation is trivial, and $a_{{\bm{k}}_i,\text{out}}=a_{{\bm{k}}_i,\text{in}}$, so that the early- and late-time vacua are the same. Indeed, for this field, the particle concept is well-defined at all times, and there is therefore a natural choice for the vacuum state~\cite{Parker1,Parker2,Parker3}. This state is called the conformal vacuum~\cite{Birrell,Parker1}. As an obvious consequence, the Bogoliubov coefficients $\beta_{\bm k}=0$ for all~$\bm k$, so no particles are created in the massless conformally coupled field case. Therefore, modes that have a positive frequency with respect to the conformal vacuum at one given time remain so for all time. Hence, if the field has no particles to begin with, expansion will not create any in the future either.

Does this mean that, if we consider a field in the conformal vacuum, no particles at all will be detected by an observer? To answer this, we must consider the fact that, while the choice of the conformal vacuum for a conformally invariant is a natural one, it is still not unique. Even in Minkowski spacetime, the Minkowski vacuum---while unique for all inertial observers---is not the vacuum for an accelerating observer~\cite{Fulling1973, Davies1975}. Such an observer would detect particles even if the field were in the ``natural'' choice for the vacuum state, i.e.,~the Minkowski vacuum~\cite{Unruh1976}. This same observer-dependence of the vacuum persists even when the usual ambiguities associated with curvature are tamed by conformal invariance. As with inertial observers in flat spacetime, however, there exists a family of privileged observers in a FLRW universe:\ those observers that see the isotropic expansion from their proper reference frame. These are called comoving observers.  The proper time of comoving observers does not coincide with the conformal time. This means that such detectors actually detect particles even in the conformal vacuum.

A particular well-known example of this is a single particle detector in a de~Sitter universe coupled to a quantum scalar field in the conformal vacuum~\cite{Birrell1982}. We will use this example to introduce a particle detector model. Consider the well-known Unruh-Dewitt detector model~\cite{DeWitt,Crispino} that describes the interaction of a two-level quantum system with a scalar field. In its simplest form, this is a pointlike particle detector whose interaction Hamiltonian with a scalar field~$\Phi(x)$ is
\begin{equation}
H_I=\Theta(\tau)\,m(\tau)\,\Phi[x(\tau)]\,,
\end{equation}
where $\tau$ is the proper time of the detector, $\Theta(\tau)$ is some appropriate switching function, $m(\tau)$ is the monopole momentum of the detector, and $\Phi[x(\tau)]$ is the field operator evaluated along the trajectory of the detector. The monopole momentum operator can be written in terms of the ladder operators $\sigma^+$ and  $\sigma^-$. We can choose to expand the field operators in terms of positive- and negative-frequency solutions of the Klein-Gordon equation in conformal coordinates $u_k(x,\eta)$ and $u_k^*(x,\eta)$, yielding
\begin{eqnarray}
\label{Hamilera}
H_{I}\!=\!g\,\Theta(\tau)\!\!\int\! \text{d} k\!\left(\sigma^+\ee^{\ii \Omega \tau}\!\!+\!\sigma^-\ee^{-\ii\omega \tau}\right)\! \Big(a^\dagger_{k} u^*_k[x(\tau),\!\eta(\tau)]\!+\!a_{k} u_k[x(\tau),\!\eta(\tau)]\Big)\,.
\end{eqnarray}
Now, considering a massless field conformally coupled to the curvature and prepared in the conformal vacuum, one can readily obtain that, at the first order in perturbation theory and for long interaction times $\Theta(\tau)\approx1$, the response function of a comoving Unruh-Dewitt detector is~\cite{Birrell}
\begin{equation}
F(E)=\frac{1}{4\pi^2}\int \text{d}\eta\int\text{d}\eta'\,\frac{\ee^{-\ii E\int_{\eta'}^\eta\sqrt{a(\eta'')}\text{d}\eta''}}{(\eta-\eta'-\ii\epsilon)^2}\,,
\end{equation}
which is nonzero in general. Hence, a comoving detector will click even if the state of the field is the conformal vacuum. A paradigmatic example of this is the de~Sitter universe, in which a comoving detector gives a thermal response~\cite{GibHawking}. The characteristic temperature of the detected particle statistics is known as the Gibbons-Hawking temperature and is proportional to the expansion rate. We will have more to say about the Gibbons-Hawking effect in Sec.~\ref{sec:confvacent}.

\section{Entanglement generation in bosonic and fermionic free fields during inflation}
\label{sec:entinflation}

As analysed in section \ref{fastexp}, when the conformal symmetry is broken there is a net particle production due to the expansion of the universe. As studied in Ref.~\cite{caball} for bosonic fields and further analysed in Ref.~\cite{fermxpanding} for both bosonic and fermionic fields, the generation of particles due to the expansion of the spacetime also generates entanglement in the final states of the field. These quantum correlations were shown to contain information about the expansion, enabling the possibility of deducing cosmological parameters of the underlying
spacetime from this entanglement. Although arguably difficult to detect in a practical cosmology experiment, this novel way of obtaining information about cosmological parameters could provide new insight into the early universe, both theoretically and experimentally. Theoretical cosmology must embrace the possibility of entanglement emerging as a purely quantum effect produced by gravitational interactions in an expanding universe and playing a fundamental role in the thermodynamic properties of FLRW spacetimes~\cite{Lousto}. Observational cosmology can now look for witnesses of purely quantum effects in the early universe. In addition to this, a parallel research effort has arisen that uses laboratory analogues of expanding spacetime to study entanglement. This is discussed further in Sec.~\ref{subsec:analogue}.

Remarkably, Ref.~\cite{fermxpanding} reveals strong qualitative differences between the bosonic and fermionic entanglement generated by such expansion. The particular way in which fermionic fields get entangled encodes more information about the underlying spacetime than that of the corresponding bosonic case, thereby allowing us to reconstruct the parameters of the history of the expansion more easily. This highlights the importance of bosonic/fermionic statistics to account for relativistic effects on the entanglement of quantum fields. This difference between fermions and bosons was already proven to be important in other relevant phenomena in relativistic quantum information, such as entanglement in non-inertial frames~\cite{AlsingSchul,Edu2,popul,beyond,Montero2011}, stationary black-holes~\cite{Kerr}, and stellar collapse scenarios~\cite{colapse}.

\subsection{Entanglement in a scalar field in a 1+1 FLRW universe}\label{scanglement}
\label{subsec:entscalar11}

Considering the 1+1 conformal metric in Eq.~\eqref{MconFLRW} with conformal factor given by Eq.~\eqref{factorbos}, we can use the solution for the Bogoliubov coefficients given by Eq.~\eqref{modolibov} directly in the formula for the `in' vacuum state, Eq.~\eqref{dynoresult}. Doing so, we find that the vacuum state of the field in the asymptotic past has the form, in the asymptotic future, of a collection of independent two-mode squeezed states~\cite{Schumaker1986} of right-moving and left-moving modes with momentum~$k$:
\begin{equation}\label{vacobos}
\ket0_{\text{in}}=\bigotimes_k\sqrt{1-\left|\frac{\beta_k}{\alpha_k}\right|^2}\,\sum_{n=0}^\infty \left(\frac{\beta_k^*}{\alpha_k^*}\right)^n\ket{n_k}_{\text{out}}\ket{n_{-k}}_{\text{out}}\,.
\end{equation}
This means that the vacuum state in the asymptotic past evolves into a mode-wise entangled state in the asymptotic future. To quantify this entanglement, we can study the entanglement entropy of the state in Eq.~\eqref{vacobos}. Entanglement entropy is only defined for bipartite pure states and is simply the von~Neumann entropy of the reduced state obtained after tracing out one of the two halves of the bipartition~\cite{Nielsen2000}. In this case, the bipartition is into left- and right-moving modes in the asymptotic future (`out' modes).

Notice that it is simply the choice of mode functions that determines whether the state appears entangled or not:\ `in' modes are separable; `out' modes are entangled. But the state is the same in both cases; it is merely being expressed in different bases. Thus, the question of whether entanglement exists in the field does not have a unique answer. This ambiguity is ubiquitous in quantum information theory:\ the presence or absence of entanglement in any quantum system can only be defined with respect to a particular choice of tensor-product structure~\cite{Zanardi2004}. In a laboratory experiment, this tensor-product structure is usually determined by asking what measurements can be performed locally. In the case considered here, it is determined by what modes are ``physically natural'' in each of the asymptotic regions. Such an ambiguity in entanglement content should not be too surprising, since we have already seen that the notion of particle counts is also non-unique, being determined by what detectors are``natural'' for a given observer.\footnote{See Eq.~\eqref{production} and the discussion above it, as well as Chapter~3 of Ref.~\cite{Birrell1982}, for more discussion and history about the concept of particles in curved spacetime.}

With respect to the `out' modes, the reduced density matrix corresponding to a right-moving mode of momentum~$k$ obtained after tracing out the left-moving modes from Eq.~\eqref{vacobos} is
\begin{equation}\label{rhoer}
\hat\rho_k=\left(1-|\theta_{\text{B}}|^2\right)\sum_{n=0}^\infty |\theta_{\text{B}}|^{2n} \ket{n_k}_{\text{out}}\!\bra{n_k}\,,
\end{equation}
where $\theta_{\text{B}}=\alpha_k/\beta_k$. Hence, the entanglement entropy~$S_{\text{B}}=-\tr(\hat\rho_k\log\hat\rho_k)$ becomes
\begin{equation}
S_{\text{B}}=\log\left(\frac{|\theta_{\text{B}}|^{2|\theta_{\text{B}}|^{2}/{(|\theta_{\text{B}}
|^{2}-1)}}}{1-|\theta_{\text{B}}|^{2}}\right)\,,
\label{Eq:entangBos}
\end{equation}
where, using Eq.~\eqref{alphak} and Eq.~\eqref{betak},
\begin{equation}
|\theta_{\text{B}}|^{2}=\frac{\sinh^2(\pi\omega_-/\rho)}{\sinh^2(\pi\omega_+/\rho)}\,.
\end{equation}
It is important to note that when the field is massless, there is no production of entanglement at all. This responds to the conformal symmetry of the theory in this particular case and is an artifact of taking 1+1 dimensions, for which minimal coupling is also conformal coupling (see Sec.~\ref{conformal}).

If the coupling to the curvature were non-zero or if we consider the massless case in higher dimensions where the conformal coupling is not equivalent to the minimal coupling, then $\beta_k$ would be non-zero, and some entanglement would be produced even in the massless case.  Although  no relevant differences would arguably be obtained in the behaviour of entanglement with the cosmological parameters, it may be useful to see how this  entanglement depends on the coupling to the curvature, thus giving insight about  fundamental properties of the underlying field theory. Although the cosmological information is codified in the generated entanglement, extracting that information---provided we could somehow experimentally access it---is not easy in the scalar case. This will be better understood after we analyse the fermionic case and we compare the amount of cosmological information codified in entanglement generated in each type of field.

In Ref.~\cite{caball}, the ability to recover information about the cosmological parameters, given access the generated entanglement, was analysed in the limit of small mass~$m\ll 2\rho\sqrt\epsilon$. They found that, in this limit, the total volume of expansion can be written as a monotonically increasing function of the entropy of entanglement
\begin{equation}
\epsilon\approx\frac{2E_k^2}{m^2}|\theta_{\text{B}}(S_{\text{B}})|\,,
\end{equation}
where $\theta_{\text{B}}=\theta_{\text{B}}(S_{\text{B}})$ is considered as a function of the entanglement entropy, inverting Eq.~\eqref{Eq:entangBos}, and $E_k=\sqrt{k^2+m^2}$ is the energy of the $k$-momentum mode. In this very same limit, the rapidity of the expansion can be related to the variation of the entanglement entropy with the energy of the studied mode:
\begin{equation}
\rho\approx\frac{\pi E}{2}\sqrt{\frac{1+|\theta_{\text{B}}(S_{\text{B}})|^2}{-\frac{E}{4}\partial_E\ln[|\theta_{\text{B}}(S_{\text{B}})|^2]-1} }\,.
\end{equation}
Unfortunately, this limit of small mass is precisely when the generated entanglement is almost zero due to the conformal symmetry of the theory. While it is true that the entanglement codifies information about the parameters of the expansion, it is not clear how to extract, in the general case, the information contained in that entanglement of scalar fields. We will revisit these results later on in section \ref{info} where we will compare these results with the fermionic case, showing that fermionic field entanglement is much better at codifying complete information about the expansion of the universe.

\subsection{Entanglement in Dirac fields in a in a 1+1 FLRW universe}\label{DiracF}

The quantisation of a spin-$\frac 1 2$ field in a curved background is somewhat similar to the process described for the scalar field in previous sections but is complicated due to the presence of spin. Let us first notate the flat-spacetime Dirac matrices as~$\{\tilde\gamma^0,\tilde\gamma^1,\tilde\gamma^2,\tilde\gamma^3\}$. They obey the anti-commutation relation~$\{\tilde\gamma^a,\tilde\gamma^b\}=2\eta^{ab}$, where $\eta^{ab}$ is the usual Minkowski metric. To study the Dirac field in curved spacetimes, we need to introduce the vierbein field~\cite{vierbein}. This field consists of an orthonormal set of four vector fields that serve as a local reference frame of the tangent Lorentzian manifold at each point of spacetime such that
\begin{equation}
g^{\mu\nu}=e^\mu_ae^\nu_b\eta^{ab}\,.
\end{equation}
The vierbein enables us to convert local Lorentz indices to general indices. With the help of the vierbein, we can write the Dirac matrices $\gamma^\mu$ in a general spacetime as a function of the local gamma matrices
\begin{equation}
\gamma^\mu=e^\mu_a\tilde\gamma^a\,.
\end{equation}
These curved-spacetime gamma matrices fulfil
\begin{equation}
	\{ \gamma^{\mu}, \gamma^{\nu} \} = 2g^{\mu\nu}\,.
\end{equation}

We need to compute the covariant derivative taking into account the fact that we are working with a spin-$\frac 1 2$ field. Since we have a spinor bundle defined over the spacetime manifold, the spin connection can be expressed in terms of the Levi-Civita connection~$\hat\Gamma^\nu_{\sigma\mu}$ as
\begin{equation}
\omega_\mu^{ab}=e^a_\nu\partial_\mu e^{\nu b}+ e^{a}_\nu e^{\sigma b}\hat\Gamma^{\nu}_{\sigma\mu}\,.
\end{equation}
We can express the covariant derivative of the fermion field as
\begin{equation}
D_\mu=\left(\partial_\mu+\Gamma_\mu\right)\,,
\end{equation}
where
\begin{equation}
\Gamma_\mu=\frac14\omega^{ab}_\mu[\tilde\gamma_{a},\tilde\gamma_b]\,.
\end{equation}
In the minimal coupling scenario, we can write the Einstein-Hilbert action for the Dirac field as
\begin{equation}
S=\int \text{d}^Dx \,\det(e)\left[\frac\ii2 e^{a\mu}\Big(\bar\psi \tilde\gamma_a D_\mu-(D_\mu\bar\psi)\tilde\gamma_a\psi\Big)\psi+m\bar\psi\psi\right]\,.
\end{equation}
This action leads to the covariant form of the Dirac equation:
\begin{equation}
(\ii e^\mu_a\tilde\gamma^aD_\mu+m)\psi=(\ii\gamma^{\mu}D_\mu+m)\psi=0\,.
\end{equation}
Let us denote its particle and antiparticle solutions of momentum~$k$ in the asymptotic past, respectively, as $u^+_{\text{in}}(k)$ and $u^-_{\text{in}}(k)$. In the same fashion, the particle and antiparticle solutions in the asymptotic future will be denoted as $u^+_{\text{out}}(k)$ and $u^-_{\text{out}}(k)$.

In the particular case of a FLRW universe, the vierbein takes the very simple form $e^{a\mu}=\sqrt{C(\eta)}\eta^{a\mu}$, where $C(\eta)=[a(\eta)]^2$ is the conformal scale factor. Again, given the  conformal symmetry (up to mass terms), we can relate the Dirac equation in the FLRW scenario to its flat-spacetime form. Using this we find, as in the scalar case, that the Bogoliubov transformation between modes in the asymptotic past and future does not mix different momentum solutions:
\begin{equation}
u_{\text{in}}^\pm(k)= \alpha_{k}^{\pm}u_{\text{out}}^\pm(k)+\beta_{k}^{\pm}u_{\text{out}}^{\mp*}(k)\,.
\end{equation}
We see that a particle mode in the asymptotic past becomes a superposition of one particle mode and one antiparticle mode in the asymptotic future.

We will now assume a specific form for the conformal factor as we did in Sec.~\ref{fastexp} so that the form of the Bogoliubov coefficients can be analytically determined.  The same form that allowed us to obtain analytical solutions in the bosonic case will not work in the fermionic case. Instead we would need the vierbein field (and not the metric as in the bosonic case) to be proportional to $1+\epsilon\tanh(\rho\eta)$. This means that if we choose a form for the conformal factor given by
\begin{equation}
C(\eta)=[1+\epsilon\tanh(\rho\eta)]^2\,,
\end{equation}
which should be compared with Eq.~\eqref{factorbos}, the FLRW metric acquires exactly the same interesting properties as in the scalar scenario. It describes a process of asymptotic flatness followed by a smooth transition to fast expansion and later followed by a smooth transition to asymptotic flatness again at a larger scale factor. The asymptotic flatness at both ends allows us to define physically meaningful `in' and `out' particle states. The similarity between this spin-$\frac 1 2$ case and the scalar case allows us to directly compare the qualitative results obtained in each.
 
Following a process completely analogous to what lead us to Eq.~\eqref{alphak} and Eq.~\eqref{betak}, we obtain the form of the Bogoliubov coefficients in the Dirac case:
\begin{eqnarray}\label{alphakF}
\alpha_k^{\pm}&=\sqrt{\frac{\omega_{\text{out}}}{\omega_{\text{in}}}}\frac{\Gamma(1-\ii\omega_{\text{in}}/\rho)\Gamma(-\ii\omega_{\text{out}}/\rho)}{\Gamma(1-\ii\omega_{+}/\rho\pm\ii m \epsilon/\rho)\Gamma(-\ii\omega_{+}/\rho\mp\ii m \epsilon/\rho)}\,, \\[2mm]
\label{betakF}\beta^\pm_k&=\sqrt{\frac{\omega_{\text{out}}}{\omega_{\text{in}}}}\frac{\Gamma([1-(\ii\omega_{\text{in}}/\rho)])\Gamma(\ii\omega_{\text{out}}/\rho)}{\Gamma(1+\ii\omega_{-}/\rho\pm\ii m \epsilon/\rho)\Gamma(\ii\omega_{-}/\rho\mp\ii m \epsilon/\rho)}\,.
\end{eqnarray}
In this case, the expression of the annihilation operators of particle and antiparticle in the asymptotic future is slightly more complicated than in the scalar case. To help, we can use the fact that the coefficients $\alpha^+_k$ and $\beta^+_k$ are not independent from $\alpha^-_k$ and $\beta^-_k$:
\begin{equation}
\frac{\alpha^+_k}{\alpha^-_k}=\frac{\omega_{\text{in}}-\mu_{\text{in}}}{\omega_{\text{out}}-\mu_{\text{out}}}\,,\qquad\frac{\beta^+_k}{\beta^-_k}=\frac{\omega_{\text{in}}-\mu_{\text{in}}}{\omega_{\text{out}}+\mu_{\text{out}}}\,,
\end{equation}
where $\mu_{\text{in}/\text{out}}=m\sqrt{C(\mp\infty)}$. As such, we obtain
\begin{eqnarray}
\label{opF1}a_{{\text{out}},\bm k,\sigma}&=
\frac{K_{{\text{in}}}}{K_{{\text{out}}}}\Bigg(\alpha_{k}^{-}a_{{\text{in}},\bm k,\sigma}+\beta_{k}^{-\ast}\sum_{\sigma'}
X_{\sigma'\sigma}(-{\bm k})b^{\dag}_{{\text{in}},-\bm k,\sigma'}\Bigg)\,, \\*
 \label{opF2}b_{{\text{out}},\bm k,\sigma}&=
\frac{K_{{\text{in}}}}{K_{{\text{out}}}}\Bigg(\alpha_{k}^{-}b_{{\text{in}},\bm k,\sigma}-\beta_{k}^{-\ast}\sum_{\sigma'}
X_{\sigma\sigma'}({\bm k})a^{\dag}_{{\text{in}},-\bm k,\sigma'}\Bigg)\,,
\end{eqnarray}
where $\sigma=\pm$. The polarisation tensor then takes the form
 \begin{equation}
X_{\sigma\sigma'}({\bm k})=-2\mu_{\text{out}}^2K^2_{{\text{out}}}\bar{U}_{{\text{out}}}(-{\bm k},\sigma')V(0,\sigma)\,,
\end{equation}
where
 \begin{equation}
K_{{\text{in}}/{\text{out}}}=\frac{1}{|\bm k|}\sqrt{\frac{\omega_{{\text{in}}/{\text{out}}}(\bm k)-\mu_{\text{in}/\text{out}}}{\mu_{\text{in}/\text{out}}}}\,,
\end{equation}
$\bar{U}_{{\text{out}}}({\bm k},\sigma)$ is a spinor solution of the Dirac equation in the curved background, and $V(\bm k,\sigma)$ is a flat-spacetime antiparticle spinor~\cite{dun1}. 
 
If we operate as we did with the scalar case and consider 1+1 dimensions, the Dirac field has no freedom in its spin and behaves as a Grasmann field~\cite{Arbitrary}. In this case, Eq.~\eqref{opF1} and Eq.~\eqref{opF2} take much simpler expressions, and the same applies to the `in' operators in terms of the `out' ones~\cite{fermxpanding}. Then, with the same technique we used to obtain Eq.~\eqref{dynoresult}, we obtain the form of the vacuum state in the asymptotic past in terms of particle and antiparticle modes in the asymptotic future demanding that the operators $a_{{\text{in}},\bm k,\sigma}$ and $b_{{\text{in}},\bm k,\sigma}$ annihilate the `in' vacuum in the `out' basis. This gives the following expression for the `in' vacuum state in terms of `out' modes:
\begin{equation}
|0\rangle_{\text{in}}=\bigotimes_{k}\frac{1}{\sqrt{1+|\theta_{\text{F}}|^{2}}}\Big(|0\rangle_{{\text{out}}}-\theta_{\text{F}}\,b^\dagger_{\text{out},k}a^\dagger_{\text{out},-k}|0\rangle_{{\text{out}}} \Big)\,,
 \label{vacin}
\end{equation}
where
\begin{equation}
\theta_{\text{F}}=\frac{\beta_k^{-*}}{\alpha_k^{-*}}\Xi(k)\qquad\text{with \ }\Xi(k)=
 \frac{\mu_{\text{out}}}{|k|}\left(1-\frac{\omega_{\text{out}}}{\mu_{\text{out}}}\right)\,.
\end{equation}
Notice that due to the Pauli exclusion principle, each mode can only be populated by a single excitation (mathematically implemented in the anticommutation relations between fermionic operators). Following identical procedures as those used in Sec.~\ref{scanglement}, the entropy of entanglement was obtained in Ref.~\cite{fermxpanding} as
\begin{equation}
S_{\text{F}}= \log\left(\frac{1+|\theta_{\text{F}}|^{2}}{|\theta_{\text{F}}|^{
\frac{2|\theta_{\text{F}}|^{2}}{|\theta_{\text{F}} |^{2}+1}}}\right)\,,
\label{SF}
\end{equation}
where $|\theta_{\text{F}}|^2$ can be simplified, using Eq.~\eqref{alphakF} and Eq.~\eqref{betakF}, to
\begin{equation}
|\theta_{\text{F}}|^2 =\frac{(\omega_- +m\epsilon)(\omega_+ +
m\epsilon)}{( \omega_-
- m\epsilon)(\omega_+ -m\epsilon)}\frac{\sinh\left[\frac{\pi}{\rho} (\omega_- -
m\epsilon)\right]\sinh\left[\frac{\pi}{\rho} (\omega_- + m\epsilon)\right]}{\sinh\left[\frac{\pi}{\rho}(\omega_+ +m\epsilon)\right]\sinh\left[\frac{\pi}{\rho}(\omega_+ -m\epsilon)\right]}\,.
\label{gamma-F}
\end{equation}
As in the scalar case, we again see that the entropy is zero when the mass of the field vanishes, this being the result of an equivalence between minimal and conformal coupling in 1+1 dimensions for Dirac fields, as well as bosonic scalar fields.

\subsection{Statistics and Information encoded in entanglement}\label{info}

\begin{figure}[t]
\begin{center}
\includegraphics[width=.60\columnwidth]{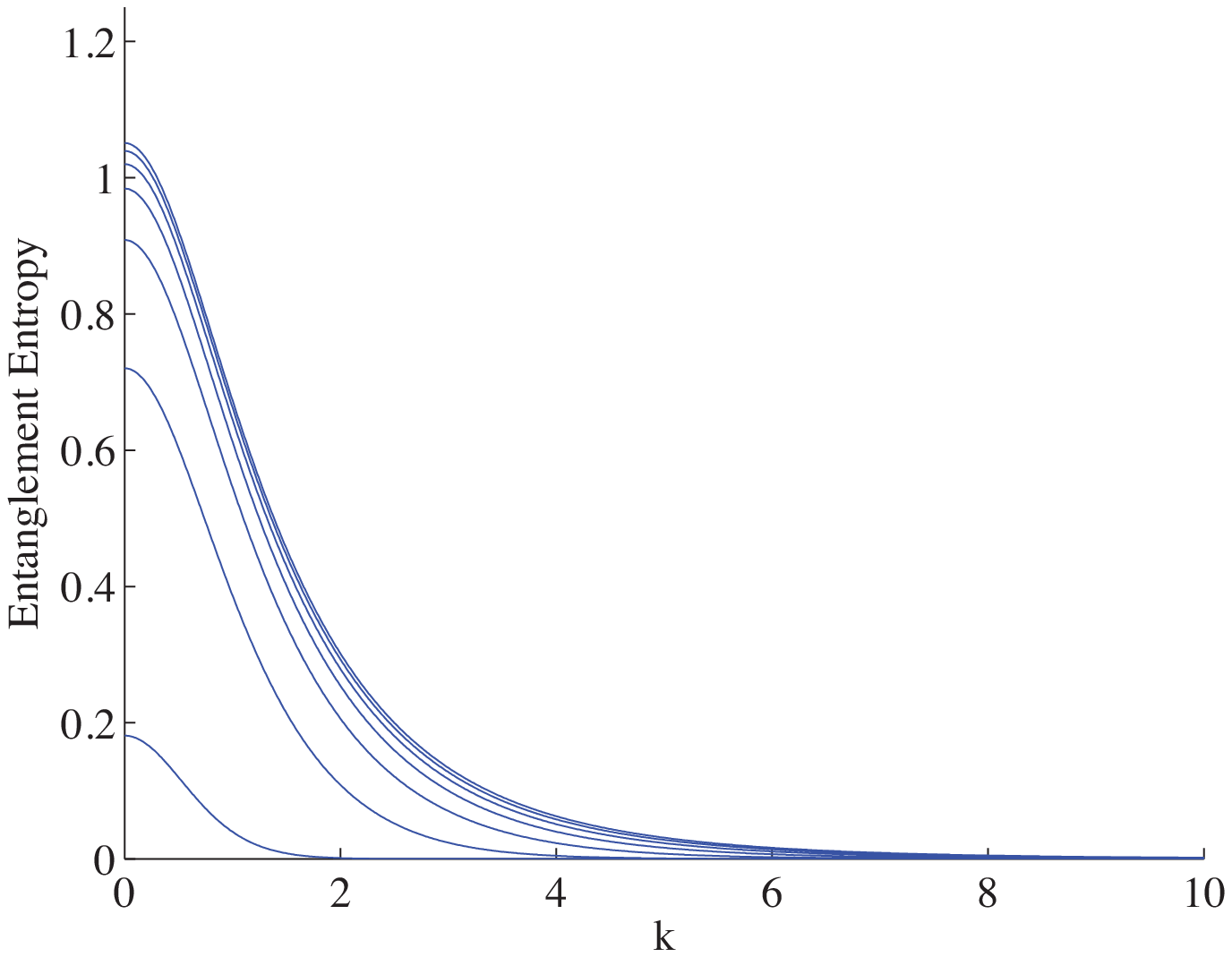}
\includegraphics[width=.60\columnwidth]{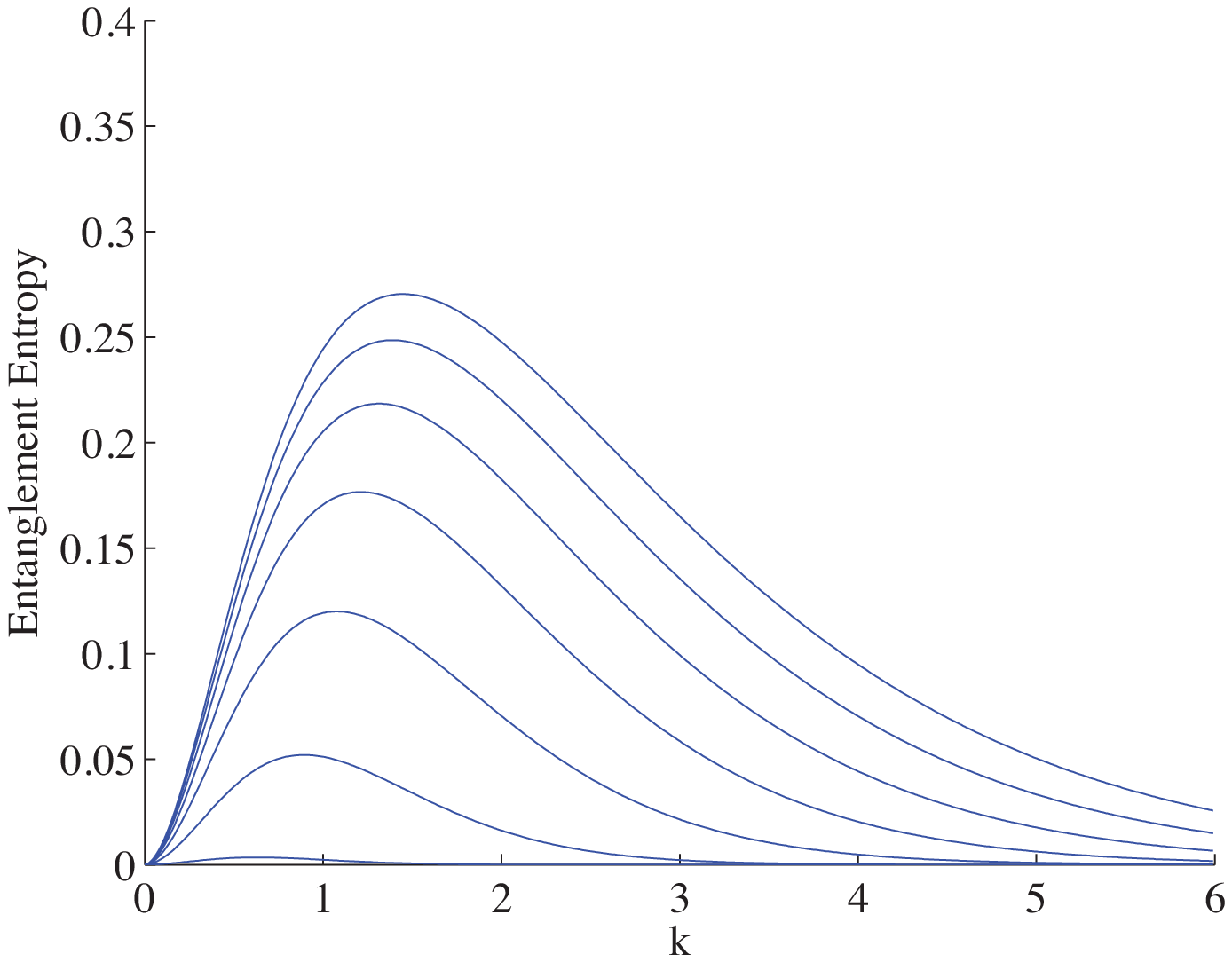}
\end{center}
\caption{{[Based on figures in Ref. \cite{fermxpanding}]} Bosonic field (top) and fermionic field (bottom): Entropy of entanglement for a fixed mass $m=1$ as a function of $|\bm k|$ for different rapidities $\rho=1,\dots,40$, with a fixed value of $\epsilon=1$.}
\label{differences}
\end{figure}

We have seen that both scalar and Dirac field modes become entangled due to the expansion of the universe when the conformal symmetry is broken. In both cases, the amount of entanglement generated in the field due to the expansion codifies information about the underlying spacetime. However, as analysed in Ref.~\cite{fermxpanding}, there are fundamental differences in the spectral behaviour of that entanglement. In Fig.~\ref{differences}, it is shown how the entanglement depends on the momentum of the field mode for different values of the parameter~$\rho$. For the Dirac case, entanglement peaks at a certain momentum, while for the scalar field, entanglement monotonically decreases with momentum and has its maximum at $|k|=0$. In other words, in contrast to the scalar field, for the Dirac field, there is a privileged value of $|k|$ for which the expansion of the spacetime generates a large amount of entanglement. Modes of that characteristic frequency are far more prone to entanglement than any others.

In Ref.~\cite{fermxpanding}, an attempt to interpret this phenomenon is given on the basis that the optimal value of $|\bm k|$ can be associated with a characteristic wavelength (proportional to $|\bm k|^{-1}$) that is increasingly correlated with a characteristic length of the universe. Intuitively, fermion modes with higher characteristic lengths are less sensitive to the underlying spacetime because the exclusion principle impedes the excitation of very long-wavelength modes---i.e.,~those whose $|\bm k|\rightarrow0$. For bosons, where this constraint does not exist, the entanglement generation is higher when $|\bm k|\rightarrow0$. This can be intuitively explained by the fact that modes of smaller $|\bm k|$ are more easily excited as the spacetime expands since it is energetically much cheaper to excite smaller $|\bm k|$ modes.

Regardless of its explanation, this natural emergence of a privileged mode in the fermion cae is a phenomenon that is very sensitive to the expansion parameters and thus more efficiently encodes information about the underlying spacetime than in the scalar case. In Ref.~\cite{fermxpanding}, it is observed that, for a given mass, the frequency for which entanglement peaks is extremely sensitive to the rapidity of the expansion of the universe, while at the same time it is almost insensitive to variations on the total volume of the universe. In contrast, the amount of entanglement in the maximally entangled mode of the field only depends on the total volume of expansion, being insensitive to the expansion rapidity. These two features, which are a direct consequence of the peaked behaviour of $S_{\text{F}}(|\bm k|,m)$, are telling us that the information about all the parameters of the expansion is codified in this peak, and if we were able to probe the field and find the most entangled mode, this will tell us much about the characteristics of the spacetime background for a fixed expansion model.

While the expansion of the universe without conformal symmetry has been proven to generate entanglement regardless of the nature of the quantum field, we have seen that Dirac fields seem to codify more information about the underlying expansion, revealing that field statistics can play a key role in the way in which the expansion of the universe generates entanglement in quantum fields.

\section{The conformal vacuum, particle detection, and entanglement}
\label{sec:confvacent}

We saw several examples above of the fact that the entanglement present within a quantum field is only loosely related to what is detected experimentally. In particular, we saw in Sec.~\ref{subsec:entscalar11} that in a conformally invariant setting, expansion produces no entanglement between the field modes and thus, supposedly, no particles. In a particular sense, as discussed in that section, this is correct. But in another sense---that of local particle detectors---this is incorrect since local detectors with a fixed resonance frequency will see a nonzero signal under expansion even under the condition of conformal invariance, as we saw in Sec.~\ref{conformal}.

One might argue that, because there is no entanglement generated by the expansion, such detectors do not see ``real'' entanglement. But entanglement is a physical resource that can be used for quantum communication and computational tasks~\cite{Nielsen2000}, so this question should have a definite empirical answer. In fact, such an answer is available:\ the entanglement is real and useful (at least in principle). In the conformal case, it is not, however, generated by the expansion; it was present in the field to begin with. The effect of expansion on the swapping of this preexisting field entanglement to local quantum systems is an interesting topic in its own right and is discussed below.

\subsection{Curvature and thermal radiation}

The fact that black holes radiate~\cite{Hawking1975} due to the quantum nature of the fields that surround them came as a shock to the physics community, overthrowing the prevailing intuition of black holes as universal vacuum cleaners (puns intended). On the other hand, the result was not entirely unexpected in light of general thermodynamic arguments about black holes being put forth at the time~\cite{Bekenstein1973}. Similar results followed for acceleration~\cite{Davies1975, Unruh1976} and---most relevant to this work---exponential expansion (i.e.,\ a de~Sitter universe)~\cite{GibHawking}, as mentioned at the end of Section~\ref{conformal}.

The groundbreaking result by Gibbons and Hawking~\cite{GibHawking} showed that an inertial particle detector in an exponentially expanding universe responds as if it were bathed in a thermal bath (in the rest frame of the detector) with a temperature proportional to the expansion rate~$\kappa$:
\begin{equation}
\label{eq:GHentpower}
	T_{\text{GH}} = \frac {\kappa} {2\pi}\,.
\end{equation}
Notice that this is the same formula as that for the temperature of Hawking radiation from a black hole~\cite{Hawking1975}, in which $\kappa$ would represent the surface gravity of the black hole. It is also the same formula as for Fulling-Davies-Unruh radiation~\cite{Fulling1973, Davies1975, Unruh1976}, in which $\kappa$ would then represent the acceleration of the detector. The common element in all three cases is the presence of a horizon, which hides the region behind from observation. As such, even if the overall state of the field is pure, it becomes mixed because of the lack of access to part of it.\footnote{This is the standard explanation, but recent results have challenged this view~\cite{Rovelli2011a}.} Black holes have an event horizon, accelerating observers have a Rindler horizon at a uniform proper distance behind them, and observers in an expanding universe perceive cosmic horizon surrounding them.

Returning our focus to Gibbons-Hawking radiation, it is interesting to note that the thermal form of the detected radiation is robust to the details of the field undergoing inflation~\cite{GibHawking}. Also of note is the fact that there is no redshift to the perceived thermal radiation spectrum. That is, the thermal radiation is always perceived as being in the rest frame of the inertial detector. This should give the reader pause when considering that two inertial detectors in relative motion will each perceive a thermal bath at rest in their respective frames, both being at the Gibbons-Hawking temperature.

\subsection{Entangling power of an expanding universe}
\label{subsec:entpower}

Consider the situation in which two detectors are both on comoving trajectories (which are themselves inertial) in a de~Sitter universe. Each detector perceives a thermal state of the field in its rest frame. But such a response would also be obtained if the spacetime were Minkowski and the field had just been heated up to an actual thermal state in some rest frame, with both detectors at rest in that same frame. As such, \emph{a single inertial particle detector cannot distinguish an empty but exponentially expanding universe from a heated one that is not expanding at all}. An interesting question then is whether entanglement between two such detectors can reveal a difference.

Reznik, Retzker, and Silman~\cite{Reznik2005} showed that entanglement in the Minkowski vacuum can be swapped to local quantum systems by coupling such systems to the quantum field for a finite time. We call such systems ``detectors'' even though the interaction with the field is coherent. Ref.~\cite{VerSteeg2009} builds on this idea by modifying it in two ways and comparing the resulting entanglement in each case. In one modification, the Minkowski vacuum of a massless scalar field is replaced with the same field in a thermal state at rest with respect to both detectors, while the metric remains Minkowski. In the other case, the Minkowski vacuum is replaced by a conformally invariant scalar field in the de~Sitter conformal vacuum, with the detectors on (inertial) comoving trajectories.

Choosing the conformal vacuum, which greatly simplifies the calculations, can be justified on physical grounds because it coincides with the massless limit of the adiabatic vacuum for de~Sitter spacetime~\cite{Birrell1982}. Thus, we can think of this analysis as applying to the following two ways of adiabatically modifying the Minkowski vacuum: (1) very slowly heating the universe to a temperature~$T$, and (2) very slowly ramping up the de~Sitter expansion rate (from zero) to a final value of~$\kappa$, with $T = \kappa/2\pi$, with the detectors activated only long after this smooth transition is complete. In both cases, each detector responds as if it were bathed in thermal radiation in its own rest frame. The temperature in the thermal case and the expansion rate in the de~Sitter case are chosen to satisfy Eq.~\eqref{eq:GHentpower}, thus ensuring identical responses. As such, access to one detector alone cannot distinguish the two cases.

\begin{figure}[tbp]
\begin{center}
\includegraphics[width= .60\columnwidth,angle=0]{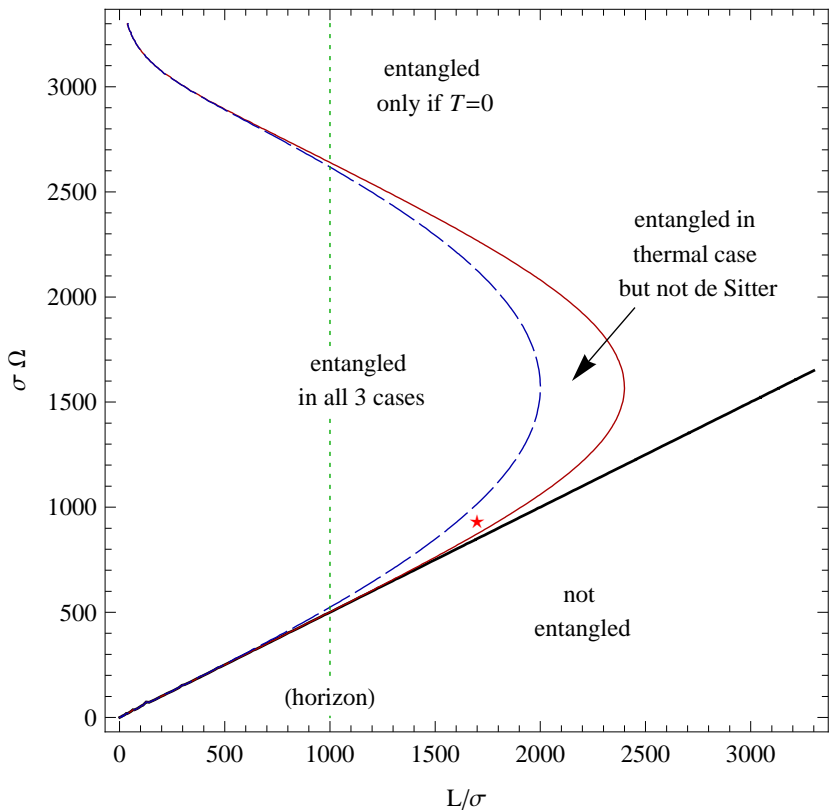}
\caption{(Based on figure in Ref.~\cite{VerSteeg2009}) Entanglement profile for detector pairs in several universes---$\sigma$ is detection time, $\Omega$ is detector resonance frequency, $L$ is detector separation.  The slanted black line is the entanglement cutoff in the Minkowski vacuum case (entangled above, separable below).  The solid red curve is the thermal Minkowski cutoff, and the dashed blue curve is the de~Sitter vacuum cutoff, both with perceived local temperatures satisfying $2 \pi T = 10^{-3} \sigma^{-1}$.  The de~Sitter horizon distance ($10^3 \sigma$) is given by the dotted green line.  The red star indicates one particular detector setup that could be used to distinguish expansion from heating.}
\label{fig:threshold}
\end{center}
\end{figure}

The coupling of the detector to the field is taken to be an Unruh-De~Witt coupling, like the one described in Section~\ref{conformal}, with a fixed resonance frequency~$\Omega$, separation in comoving coordinates~$L$, and time-dependent coupling strength taken to be a small fixed parameter times a Gaussian in (proper) time with standard deviation~$\sigma \ll L$. While the detector technically remains coupled to the field for all times due to the infinite tails of the Gaussian, a smooth cutoff applied in the wings of the Gaussian was found not to change the results, so the full Gaussian form is used for calculational simplicity to approximate a coupling that is on for a time~$\sigma$. The goal is to find a set of parameters~$(\Omega, \sigma, L)$ such that the detectors (two-level systems) would become entangled through the interaction in one case (heating or expansion) but not the other.

One might expect that the entanglement persists---or even grows---in the case of de~Sitter expansion (since the full state remains pure), while it gets drowned out by actual particles in the corresponding heated case. Counterintuitively, the opposite occurs. The results are shown in Figure~\ref{fig:threshold}. There is a region of parameter space in which a heated universe entangles the detectors while an expanding universe at the same perceived temperature does not. As such, the two types of universes can be distinguished by their ``entangling power'' with respect to local inertial detectors, although perhaps not in the way one would guess.

For it to have operational meaning, the claim that entanglement can be used to distinguish expansion from heating must be capable of being verified by observers within the universe. One thing to notice right away is that if the universe is expanding, there is a cosmic horizon distance beyond which two parties cannot communicate~\cite{GibHawking}. Distinguishing heating from expansion requires that the detectors be separated by more than this distance. Thus, they cannot communicate with each other to verify their entanglement, nor can they use the entanglement for quantum teleportation or long-distance quantum communication with each other, since teleportation still requires classical communication of local measurement results~\cite{Bennett1993}. However, if we assume that the detectors are themselves mounted on satellites and that each satellite can communicate with a ``home planet'' located half-way in between them, results of local measurements confirming violation of a Bell inequality can always, in principle, be transmitted to a third party on the home planet for verification~\cite{Masanes2006,Horodecki1997}. In addition, given an initial source of entanglement between the home planet and each satellite separately, the entanglement extracted from the quantum field could be swapped via teleportation back to the home planet~\cite{Bennett1993, Zukowski1993}. This method of remote ``entanglement harvesting'' from a quantum field is nowhere near practical since the effect is unfathomably small~\cite{VerSteeg2009}, but the thought experiment shows that in principle the existence of this entanglement is verifiable since it can be used as a quantum information processing resource back on the home planet.

Several results are notable here. First, in both the thermal Minkowski and de~Sitter vacuum cases, the region of entanglement is a proper subset of that of the Minkowski vacuum case. In the thermal case, this is expected because the state is now more mixed, but this is somewhat surprising for the expanding case. The results in Section~\ref{subsec:entscalar11}, however, gives a possible explanation:\ no entanglement is generated by the expansion in a conformally invariant system. While this is not a rigorous explanation, it accords with the intuition that the radiation detected should not be any more useful (in terms of generating the physical resource of entanglement) than the vacuum itself.

The reason that entanglement is lost in the expanding case can be understood by considering the entire system in conformal time instead of in the detectors' proper time. Since the field is conformally invariant, the modes behave like ordinary Minkowski modes with respect to conformal time. Therefore, detectors with a fixed resonant frequency in conformal time in an expanding universe will see the same entanglement profile as that for the Minkowski vacuum because such detectors are tuned to a single field mode. Having a fixed frequency in conformal time necessarily means having a time-dependent frequency in detector proper time (for an inertial detector), and thus, a fixed frequency in proper time means coupling time-dependently to many different independent field modes in an expanding universe. This increases local noise and reduces entanglement. This method of exchanging one type of detector behavior in conformal time with another type in proper time was recently proposed as a way to implement the ion-trap analogue of a detector in an expanding universe~\cite{Menicucci2010a}. This proposal and several other possible analogue models are compared in Section~\ref{subsec:analogue}.

\section{Entanglement and quantum correlations in the cosmic background}

\subsection{Inflation and the seeds of cosmic structure}

The inflationary hypothesis solves numerous problems with standard cosmology, including its two original targets:\ the horizon problem and the flatness problem~\cite{Guth1981}. The horizon problem is the observation that the universe would not have had time to organize its own large-scale homogeneity through causal means in the time available since the Big Bang. The flatness problem is the observation that the spatial curvature of the universe appears to be fine tuned such that it exists for $\sim 10^{10}$~years in a metastable state instead of collapsing in on itself or expanding to complete emptiness on the scale of the Planck time ($5.4 \times 10^{-44}$~s).

Both problems are solved by postulating that the highly disordered, volatile, ``curvy and bumpy'' universe immediately after the Big Bang undergoes a period of rapid expansion, flattening out any curvature and flinging all matter and energy beyond a cosmic horizon~\cite{Guth1981}. The scale factor during this period is exponential in the cosmic time and thus approximates a de~Sitter universe. But as we know from quantum field theory, what is left after all matter and energy are gone---empty space---is not entirely empty. As discussed in Section~\ref{sec:confvacent}, de~Sitter expansion will cause the observable portion of the universe to appear nonempty (thermal) despite the entirety of it---or at least a much larger portion of it---being in a vacuum state. As such, the expansion causes the universe to possess tiny fluctuations (about a part in $10^5$) in an otherwise perfectly smooth background. 

These fluctuations are taken to be the seeds of cosmic structure~\cite{Genovese2009}. The colloquial explanation for this is that quantum fluctuations in the modes of the inflaton field (the scalar field responsible for inflation) become ``frozen'' when the size of the cosmic horizon (the inverse of the Hubble parameter) becomes comparable to the wavelength of each mode, creating tiny fluctuations in the background gravitational field, thus providing the necessary nucleation points for galaxies and other cosmic structures. Therefore, inflation now also provides a means to understand structure formation.

The main evidence for inflation comes from the cosmic microwave background~(CMB), a nearly uniform thermal glow of the universe at a temperature of 2.725~K, first discovered by accident~\cite{Penzias1965} and later studied in detail by the sky mapping experiment known as the Cosmic Background Explorer~(COBE). In addition to revealing a nearly perfect blackbody spectrum~\cite{Mather1990}, COBE also revealed tiny fluctuations in temperature---signatures of the tiny fluctuations that led to structure formation in the early universe~\cite{Smoot1992}. Its resolution was too low, however, to compare with specific inflationary models. Mapping the details of the anisotropy of the CMB was a task undertaken by the Wilkinson Microwave Anisotropy Probe~(WMAP), which produced images with 30 times the resolution of those from COBE~\cite{Bennett2003} and enabled constraints to be placed on inflationary models based on comparison of their predictions of structure formation with the reality of fluctuations in the CMB~\cite{Peiris2003}.

\subsection{Cosmic fluctuations: the quantum-to-classical transition}

While the inflationary hypothesis, broadly speaking, solves the horizon and flatness problems, its proposal for solving the problem of structure formation leaves us wanting for details:\ what specifically is the physical mechanism by which \emph{quantum} fluctuations in a scalar field are ``frozen'' by inflation into eventual \emph{classical} density fluctuations in the matter distribution of the post-inflationary universe (and, eventually, temperature variations in the CMB)? We would like to have a better understanding of this process using the rigorous tools of quantum information theory.

Several avenues for pursuing this question are available~\cite{Genovese2009}, and the answers given depend on the model in question and possibly even on one's interpretation of quantum mechanics. Some authors~\cite{Perez2006} go so far as to postulate a need for new physics to explain the transition, with an appeal to dynamic collapse models---in which gravity or some other principle induces a real, unambiguous, dynamical collapse of a superposition into one of its constituent states. We don't need to go this far, however, to say something meaningful about the quantum-to-classical transition needed to account for cosmic structure formation. By studying the entanglement between expanding modes or regions of space and by considering the effects of decoherence caused by interactions within the inflaton field and with other fields, we can begin to discuss this physical process meaningfully.

The approach by Nambu~\cite{Nambu2008} is firmly rooted in the methods of Gaussian quantum information science and employs a one-dimensional lattice model of a scalar field during inflation. The lattice is separated into two finite, continguous, nonadjacent, nonoverlapping regions, and the entanglement between them is monitored as inflation progresses. Because the initial state is Gaussian and the field has no nonlinear terms, it remains Gaussian throughout the inflation, and the criterion of positive partial transpose for Gaussian states can be used to signal decisively the presence or absence of entanglement~\cite{Simon2000}. The results indicate that the two regions becomes disentangled when the size of the regions exceeds Hubble horizon distance. This is claimed as a necessary condition for the appearance of classicality of the fluctuations by virtue of entangled states failing to admit local hidden variable models for their statistics~\cite{Bell1964}. The model is interesting because it is straightforward and simple, using only Gaussian states, and is therefore amenable to direct analysis of its entanglement structure. But this is also a weakness in that non-Gaussian features of inflation are not captured in this picture~\cite{Genovese2009}. Furthermore, it focuses on the emergence of classicality within the field itself, and as we have seen above, properties that hold for field modes may not translate directly to what a detector (of any sort) would see.

The effect of expansion on local detectors has been discussed in Section~\ref{subsec:entpower}, but the results used a conformally coupled scalar field, instead of a minimally coupled field, which which is usually assumed in inflationary models. The results of Section~\ref{subsec:entscalar11} indicate that conformal coupling is quite special, and they suggest that minimal coupling will produce a very different result. Nambu and Ohsumi~\cite{Nambu2011} address this and generalise the analysis presented in Ref.~\cite{VerSteeg2009} to a minimally coupled scalar field. This analysis involves particle detectors, thus providing a stronger grounding in operationalism than Ref.~\cite{Nambu2008}, but the basic goal of understanding is the same:\ to analyze the entanglement properties of a quantum field during inflation, either directly in the field itself or by considering the response of local detectors. The results of this study for minimally coupled fields show that entanglement vanishes at the Hubble horizon distance, in contrast to the result for conformal coupling~\cite{VerSteeg2009}, in which it vanishes at about twice the horizon distance. Nevertheless, the quantum discord~\cite{Ollivier2001, Vedral2003, Luo2009} in the detectors---a measure of nonclassicality that may be present even without entanglement---remains even when separated beyond the horizon but eventually goes to zero for sufficiently separated detectors. These results show that conformally coupled and minimally coupled scalar fields show different behavior in terms of their entangling power.

The search for signatures of inseparability in quantum fields during inflation includes a study by Campo and Parentani~\cite{Campo2005}. The analysis proceeds by modeling the effects of inflation on individual field modes as a two-mode squeezing operation~\cite{Schumaker1986}. Decoherence is then added to simulate the average effect of small nonlinearities in the coupling that would otherwise cause the evolution to deviate from a Gaussian unitarity. The results are, unsurprisingly, that the two-mode squeezing operation produces entanglement between modes of opposite wave vector and that under sufficiently low decoherence, these states will always violate some Bell inequality~\cite{Bell1964}. Furthermore, there is a level of decoherence above which no Bell inequality is violated. Practical detection of this violation is a near impossibility, but the study nevertheless shows that if decoherence induced by nonlinearities in the evolution is small enough, the quantum nature of the fluctuations should survive.

The question of nonlinearities in the evolution was taken up directly by a simulation reported by Mazur and Heyl~\cite{Mazur2009}. Instead of just two modes, $\pm \mathbf{k}$, four modes are used: $\pm \mathbf{k}$ and $\pm 2 \mathbf{k}$. Terms are added to the field Lagrangian that induce couplings of the form $g a_{\pm 2\mathbf{k}} a^\dag_{\pm \mathbf{k}} a^\dag_{\pm \mathbf{k}}$ and $g^* a^\dag_{\pm 2\mathbf{k}} a_{\pm \mathbf{k}} a_{\pm \mathbf{k}}$, which are nonlinear and allow particles to be exchanged:\ two particles from mode~$\pm \mathbf{k}$ for one particle in mode~$\pm 2\mathbf{k}$ and vice versa. It is then assumed that coarse-graining of measurements only allows the larger modes ($\pm \mathbf{k}$) to be detected, requiring the smaller modes at twice this wave vector to be traced out, thus generating entropy. Once again, entropy generation---and thus randomness in a measurement outcome---is taken as a sign of classicality. The results show that nonlinear interactions cause decoherence in individual modes by allowing the swapping of quanta to modes of higher or lower energy. These results show that reheating (post-inflationary particle production) is not the only mechanism for decoherence, as is commonly assumed. Nevertheless, the decoherence predicted is quite small and is not enough on its own to explain classicality for fields that do not undergo reheating.

These results point to an interesting question, which is addressed by Franco and Calzetta~\cite{Franco2011}. The question comes from applying scalar and tensor perturbations to a spacetime metric. Scalar perturbations in the metric at the end of inflation show up as temperature variations in the CMB, while tensor perturbations are only detectable in the CMB polarisation~\cite{Kaplan2003}. Therefore, if our instruments are only measuring temperature variation (not polarisation), then the tensor perturbations act as an environment for the scalar ones, causing decoherence. The authors use the Gell-Mann--Hartle histories formulation of quantum mechanics~\cite{Gell-Mann1993} to show that this is indeed the case.

There remains no consensus on the mechanism by which quantum fluctuations become classical density fluctuations during (or after) inflation. While studying entanglement and decoherence in the field allows us to at least approach the problem, the interpretation of such results must be done carefully because it involves, at its heart, one of the most controversial aspects of quantum theory: the measurement problem.

\subsection{Relation to the measurement problem in quantum theory}

Any discussion of the quantum-to-classical transition, whether in a table-top laboratory experiment or on cosmic scales, necessarily begins to wade into interpretations of quantum mechanics and metaphysics, and thus we must tread carefully to ensure we are actually discussing that which we believe we are discussing. For instance, no one worries that a linear polarisation measurement made on a circularly polarised photon reveals a definite, classical answer even though the outcome that can only be predicted stochastically. Even if the photon is entangled with a distant one, we still get a definite, classical answer. ``Collapse'' is a part of everyday life in experimental physics.

As such, one should not expect that the temperature map of the CMB obtained by COBE or WMAP would ever appear to be in a superposition of possible maps. If inflation predicts a certain power spectrum for temperature fluctuations of various sizes in the CMB, then we would expect that the fixed, definite, classical temperature map revealed by WMAP would be statistically \emph{typical} if drawn from that distribution of theoretical possibilities. This is the nature of all stochastic prediction. We stress this because the usual story is that it is entanglement in the \emph{inflaton} field---not in the electromagnetic field, which is the field being directly observed by the satellite experiments---that is responsible for the density fluctuations in the early universe, which then, in turn, become imprinted on the CMB in the form of measurable temperature fluctuations. It is not fruitful, therefore, to ponder why our instruments detect a definite temperature map instead of a superposition of such maps:\ once a measurement is made, superpositions between the possible outcome states are destroyed, and a definite answer is obtained. Such is the case with our mapping of the CMB temperature.

On the other hand, if the primordial entanglement from the inflaton field were somehow transferred to other fields that we have access to today, such as the electromagnetic field, then it is not impossible (at least in principle) that we might observe entanglement in the individual photons of the CMB coming from different spatial directions by looking at correlation statistics over many observations, such as is done to verify entanglement in parametric down-conversion experiments~\cite{PDC,Aspect}.

This is exceedingly unlikely, however, at least for the electromagnetic field, because entanglement and quantum coherence are extremely fragile. We do not need to look far for things that destroy them. Decoherence is a constant nuisance in experimental quantum physics, and as such, the burden of proof rests squarely on the shoulders of those who wish to claim that long-range coherence could be maintained for millennia after inflation ended. Even if individual field modes are entangled as a result of the gravitational interaction, once they interact with anything else in the universe, their coherence will quickly be lost unless the fields are exceptionally weakly interacting. We fully expect there are a number of such decoherence mechanisms through self-coupling and coupling to other fields, including gravity. It is important to determine which of these processes are dominant, and it is here that the community may fruitfully focus its efforts.

\section{Prospects: Quantum entanglement as a tool for Cosmology}

In the hopes of stimulating further research into the connection between quantum entanglement and cosmology, we conclude with some discussion of further research avenues in this area. We discuss the possibility of the cosmic neutrino background possessing entanglement from the early universe and surviving all the way to the present due to its weak coupling to matter. We also discuss the importance of analogue experimental models that provide a laboratory-accessible testbed for cosmological models. Finally, we conclude with a brief summary of possible future prospects for using quantum entanglement as a tool for theoretical and experimental cosmology. 

\subsection{Quantum correlations in the cosmic neutrino background}
\label{subsec:neutrinoback}

We have seen in previous sections that the quantum correlations present in quantum fields in the early universe would be hard-pressed to survive to the present due to decoherence induced by field interactions. This means that if remnants of these quantum correlations are to survive and be detectable today, they must do so in those quantum fields that interact very weakly. In this regard, the CMB is highly non-optimal since the electromagnetic field interacts with everything that has electric charge, thus leaving a background that is almost certainty decohered by now. Alternatively, the cosmic neutrino background~\cite{CvB} does not have this problem since neutrinos rarely interact with anything at all. As such, the neutrino background might prove to be a better candidate for residual quantum correlations than the CMB.

Due to the fact that these low energy neutrinos only interact weakly with the rest of the quantum fields they are extremely difficult to detect so that, unlike the CMB, the neutrino background is not easily directly observable. There is, however, indirect experimental evidence of its existence~\cite{EvidenceCvB}, and it is estimated that today the neutrino background has a temperature of roughly 2~K. On the other hand, the neutrinos' introverted nature, which thwarts our efforts to detect them, nevertheless becomes a feature when it comes to the study of quantum correlations in the primitive universe, since the strong quantum correlations present in the early universe would have been protected from decoherence by a lack of interaction with other fields.

Despite the difficulty of its detection, if our sole criteria is to improve on the CMB as a possible source of primordial cosmic entanglement, then the neutrino background remains a candidate. Searching for entanglement in the CMB is severely hindered by the fact that the light we observe today dates back only to recombination---the formation of the first stable atoms and the resulting decoupling of matter and radiation that followed---which occurred approximately $\sim10^5$ years after inflation was to have ended. That's an impossibly long time for entanglement to survive under strong coupling to matter. Neutrinos decoupled from matter much earlier than the electromagnetic field---approximately one second after inflation~\cite{Longair2006}. As such, the required coherence time for any quantum entanglement to survive in the neutrino background (assuming no decoherence during its travel to our detectors) is 12 orders of magnitude lass than what would be required for entanglement to survive in the CMB (one second versus $10^5$~years). Despite the fact that the idea of probing the neutrino background for primordial cosmic entanglement is nowhere near a practical proposal, these arguments suggest we should not dismiss the possibility out of hand.

\subsection{Analogue models}
\label{subsec:analogue}

 Most predictions of curved-spacetime QFT, including nearly all of the effects discussed in this paper, are extremely difficult to test directly, confined as we are to a low-curvature region of spacetime, with dimensional analysis conspiring to make the most straightforward effects extremely tiny.  This situation has given rise to a number of proposals for testing \emph{analogue} quantum field theory in curved spacetime, the seminal example being Unruh's proposal for detection of the sonic analogue of Hawking radiation in supersonic fluid flow~\cite{Unruh1981}.

In such experiments, spacetime is replaced by a system of coupled atoms, either as discrete entities or in a fluid, and field modes are replaced by the collective normal modes of oscillation of the atoms---effectively a ``phonon field''~\cite{Barcelo2005}. The advantage of this over probing fundamental quantum fields in outer space is that, while the observable effects in the laboratory setup are conceptually the same as (analogous to) those predicted for quantum fields on a curved background, the parameters of the experiment may be adjusted to give much stronger and more easily observed effects. The most common analogue curved-spacetime field theory proposals involve Bose-Einstein condensates~(BECs)~\cite{Fedichev2003, Fedichev2004, Uhlmann2005, analog2} or ion traps~\cite{Schutzhold2007, Menicucci2010a} as the analogue spacetime. We say a few words about these here and refer the reader to the aforementioned references, plus the review articles of Refs.~\cite{Schutzhold2009, Barcelo2005}, for more details. We also note that liquid helium has also been proposed as an analogue testbed~\cite{Volovik}, as well as schemes based on photons in a microwave guide and superconducting circuits \cite{DelRey2012}.

BECs provide a natural system for studying the analogue of FLRW expansion~\cite{Fedichev2003, Fedichev2004, Uhlmann2005, analog2}. The condensate field is linearized into a classical background plus a quantum field of small fluctuations that lives on top of it. When a fully condensed system is allowed to expand in the laboratory, the emulated effect is the expansion of the spacetime on which the quantum fluctuations live. Furthermore, as expansion proceeds in the laboratory, the initial correlations in the ground state of the field naturally evolve into classical fluctuations in the density of the background condensate, which can be observed simply by taking a picture of the condensate during expansion. In a freely expanding BEC, a sonic horizon is formed, allowing an experimental study of the ``freezing'' of quantum fluctuations into a classical density distribution as the size of a given mode stretches beyond the sonic horizon.

The main advantage of trapped-ion analogue gravity experiments~\cite{Schutzhold2007, Menicucci2010a} over other proposals, such as BECs, lies in the exquisite quantum control over state preparation, interaction, and readout~\cite{James1998}. Each ion is equipped with its own motional detector in the form of laser-induced coupling of its vibrational motion to its electronic state. Effectively, then, we have a discrete phonon field with a detector mounted at every point that we can turn on and off at will in a continuous and highly controlled manner. In addition, cooling to the ground state is achievable with high fidelity. The tradeoff for this precision and ``digital'' nature of the field and its detectors is in the small maximum number of ions confinable to a single linear trap (currently, about ten~\cite{James1998}). Two methods of using ion traps to study cosmic expansion have been proposed. The first~\cite{Schutzhold2007} allows the ions to physically expand for a time, simulating expansion starting and then stopping, as in Sec.~\ref{fastexp}. Another proposal~\cite{Menicucci2010a} uses the properties of conformally invariant expansion to encode all details of the scale factor in the parameters used to couple the electronic and vibrational degrees of freedom. This allows detector responses during expansion of a conformally coupled field to be emulated in the ion trap without physically moving this ions at all. (Similar techniques are used in BECs, as well~\cite{Fedichev2003, Fedichev2004}.) In addition to these applications, a linear ion trap can be used to emulate entanglement swapping from a flat-spacetime quantum field to local detectors~\cite{Retzker2005}. Ref.~\cite{Menicucci2010a} proposed modifying this setup to perform the analogue of the thought experiment proposed in Ref.~\cite{VerSteeg2009} and discussed in Sec.~\ref{subsec:entpower}, possibly bringing entanglement-based discrimination of (analogue) gravitational parameters into the realm of laboratory testability.

\subsection{Other research avenues}

\label{subsec:future}

We end with some thoughts on further research avenues in this area.

\emph{Coupling to curvature}---An important observation is that conformal invariance, while greatly simplifying calculations for expanding spacetimes, is often not present in inflationary models. Minimal coupling is used instead. While the two settings coincide in two dimensions, they differ in every other dimension, including that of our own \text{4-dimensional} universe. There are substantial differences between the physics of the two types of coupling, even in the massless case, as we have stressed many times throughout this article, with the most important being that expansion of a conformally coupled field generates no entanglement in the field modes and no particle production if the expansion ceases (Sec.~\ref{subsec:entscalar11}). Therefore, analyses that include the ability to select the conformal coupling constant (or at least to allow it to be set to zero) are a natural generalisation of many of the results presented here and would allow us to examine how the entanglement produced scales with this coupling strength.

\emph{Entanglement extraction}---In addition to the entanglement produced in the field modes themselves, we can also ask how the field coupling to curvature affects entanglement extraction from the field using local detector couplings. Ref.~\cite{Nambu2011} pursued this as a first step, generalizing the results of Ref.~\cite{VerSteeg2009} from conformal to minimal coupling. But we know from analogue proposals~\cite{Fedichev2003, Fedichev2004, Menicucci2010a}, as well as from more recent results showing that entanglement exists between~\cite{Olson2011} and can be extracted from~\cite{Olson2012} the past and future wedges of a spacetime (instead of from the left and right wedges), that modulating a detector's resonance frequency and interaction strength can be used to optimize the extracted entanglement. Even the early proposal for this procedure, Ref.~\cite{Reznik2005}, introduced a rather complicated time-varying coupling to amplify the amount of entanglement extracted from the vacuum state. In addition to optimising the extracted entanglement, we might also ask what time dependence of the interaction would optimise the extraction of information about cosmological parameters and spacetime structure using this method.

\emph{Anisotropy of inflation}---Previous literature analysing the emergence of entanglement in quantum fields during inflationary periods presented results based on the assumption of isotropy in the expansion. The question remains open as to whether the entanglement generated during inflationary periods may be sensitive to possible anisotropies in this expansion. It is known that particle creation feels the difference~\cite{Nesteruk1991}, but it still remains to be explored to what extent this happens and how much information about the anisotropies would be encoded into the entanglement generated in the inflationary period.

\emph{Analogue models}---Perhaps one of the most promising avenues of research is in quantum emulation of expansion through analogue experiments. In the ion trap architecture, the approach of Ref.~\cite{Menicucci2010a} is complementary to that of Ref.~\cite{Schutzhold2007} in that the former modifies detector parameters to emulate expansion of a conformally coupled field, while the latter literally expands the trap, allowing the ions to manifest actual phonon production with expansion over a finite time. Perhaps the two could be combined to study more complicated phenomena. Alternatively, Ref.~\cite{Fischer2004} proposes a BEC model in which atoms in two different hyperfine states are trapped together and allowed to emulate more complex inflationary scenarios. The development of a reliable, robust, and highly tunable laboratory testbed for analogue inflation would be of great experimental value in testing predictions in theoretical cosmology.

\emph{Observational astronomy}---Our best data sets today regarding the early universe come from observations of the CMB~\cite{Bennett2003}. The possibility of using the cosmic neutrino background to search for primordial entanglement was already discussed in Sec.~\ref{subsec:neutrinoback}. Since this proposal remains far beyond current technology, the CMB is our best hope for testing theories of cosmology---through the temperature maps produced by WMAP~\cite{Peiris2003} and through ongoing efforts to map the polarisation of the CMB, which encodes cosmic information not found in the anisotropies alone~\cite{Kaplan2003}.

The research effort into the connection between cosmology and entanglement is vast, ranging from studies of entanglement in field modes, to the ability to swap this entanglement to local detectors, to the extraction of cosmological parameters from such measurements, to direct comparison of observational data and theoretical models, to the experimental probing of theoretical proposals in analogue experiments. With multiple fronts opening into unexplored territory, we hope the reader will find the results illuminating and inspiring of further research into the connection between cosmology and entanglement at the theoretical, observational, and laboratory-experimental levels.

\subsection*{Acknowledgments}

We thank Carl Caves, Achim Kempf and Miguel Montero for inspiring discussions.

\vspace{1em}
\hrule

\end{document}